\documentclass[aps,prd,groupedaddress,nofootinbib,english,nobalancelastpage,floatfix,superscriptaddress,preprintnumbers]{revtex4-1}
\pdfoutput=1
\usepackage{flushend}
\DeclareMathAlphabet{\scr}{U}{rsfs}{m}{n}
\usepackage{graphicx}
\usepackage{amsmath,bm}
\usepackage{multirow}
\usepackage{rotating}
\usepackage{float}
\usepackage[normalem]{ulem}

\usepackage{color}
\definecolor{naviBlue}{RGB}{0,0,128}
\usepackage[colorlinks  = true,
            linkcolor   = naviBlue,
            urlcolor    = naviBlue,
            citecolor   = naviBlue,
            anchorcolor = naviBlue]{hyperref}

\newcommand{\newc}{\newcommand}
\newc{\sigmav}{\langle\sigma v \rangle}

\newcommand{\eVdist}{\kern-0.06em}

\newcommand{\ie}{\emph{i.e.}}

\newcommand{\cf}{\emph{cf.}}
\newcommand{\diff}{\mathrm{d}}

\begin{document}

\title{Implications of Lithium to Oxygen AMS-02 spectra on our understanding of cosmic-ray diffusion}

\author{Michael Korsmeier}
\email{michael.korsmeier@fysik.su.se}
\affiliation{The Oskar Klein Centre, Department of Physics, Stockholm University, AlbaNova, SE-10691 Stockholm, Sweden}
\affiliation{Institute for Theoretical Particle Physics and Cosmology, RWTH Aachen University, Sommerfeldstr.\ 16, 52056 Aachen, Germany}
\affiliation{Dipartimento di Fisica, Universit\`a di Torino, Via P. Giuria 1, 10125 Torino, Italy}
\affiliation{Istituto Nazionale di Fisica Nucleare, Sezione di Torino, Via P. Giuria 1, 10125 Torino, Italy}
\author{Alessandro Cuoco}
\email{alessandro.cuoco@unito.it}
\affiliation{Dipartimento di Fisica, Universit\`a di Torino, Via P. Giuria 1, 10125 Torino, Italy}
\affiliation{Istituto Nazionale di Fisica Nucleare, Sezione di Torino, Via P. Giuria 1, 10125 Torino, Italy}

\preprint{TTK-21-10}

\begin{abstract}
We analyze recent AMS-02 comic-ray measurements of Lithium, Beryllium,  Boron,  Carbon, Nitrogen and Oxygen.
The emphasis of the analysis is on systematic uncertainties related to propagation and nuclear cross sections.
To investigate the uncertainties in the propagation scenario,
we consider five different  frameworks, differing with respect to the inclusion of a diffusion break at a few GV, the presence 
of reacceleration, and the presence of a break in the injection spectra of primaries. 
For each framework we fit the diffusion equations of cosmic rays and explore the 
parameter space with Monte Carlo methods.
At the same time, the impact of the uncertainties in the nuclear production cross sections of secondaries is explicitly considered and included  
in the fit through the use of nuisance parameters.
We find that all the considered frameworks are able to provide a good fit. In particular, two competing scenarios,
one including a break in diffusion but no reacceleration and the other with reacceleration but no break in diffusion are both allowed.
The inclusion of cross-section uncertainties is, however, crucial, to this result. Thus, at the moment, these uncertainties represent 
a fundamental systematic preventing a deeper understanding of the properties of CR propagation.
We find, nonetheless, that the slope of diffusion at intermediate rigidities is robustly constrained in the range $\delta\simeq0.45-0.5$ in models without convection, or 
$\delta\simeq0.4-0.5$ if convection is included in the fit.
Furthermore, we find that the use of the AMS-02 Beryllium data provides 
a lower limit on the vertical size of the Galactic propagation halo of $z_\mathrm{h}\gtrsim3$ kpc.
\end{abstract}

\maketitle

\section{Introduction}

Over the last decade AMS-02 has acquired precise measurements of cosmic-ray electrons and positrons 
as well as multiple nuclei from protons to iron \cite{Aguilar:2021tos}. These data are provided with unprecedented precision 
at the level of a few percent in a large energy range between 1~GeV and a few TeV and have triggered numerous 
studies \cite{Korsmeier:2016kha,Tomassetti:2017hbe,Liu:2018ujp,Genolini:2019ewc,Weinrich:2020cmw,Weinrich:2020ftb,Evoli:2019wwu,Evoli:2019iih,Boschini:2018baj,Boschini:2019gow,Luque:2021nxb,Luque:2021joz,Schroer:2021ojh}. 
The propagation of cosmic rays (CR) in the Galaxy can be modeled as a diffusive process within a halo 
extending a few kpc above and below the Galactic plane. 
While electrons and positrons lose energy quickly and can thus be used to study the local Galactic environment
(up to few kpc) CR nuclei sample a much larger volume of the diffusion halo up to about $10$~kpc. Beside  diffusion 
and energy losses, other ingredients like convective winds or reacceleration might be important.
These processes can be constrained by using secondary CRs like Boron (B), Beryllium (Be) and Lithium (Li), 
i.e., CRs produced by the fragmentation of primary CRs, mainly Carbon (C), Nitrogen (N) and Oxygen (O), on the interstellar medium (ISM) 
during propagation\footnote{The N flux contains a significant secondary component.}. 
Historically, B has been the most investigated secondary nucleus. Nonetheless, AMS-02 has now provided excellent measurements 
of Li  and Be \cite{Aguilar:2018njt}, which have been indeed studied in recent analyses. 
Beryllium is particularly important because its isotope $^{10}\mathrm{Be}$ has a life time of $\sim$ 1 Myr, similar to the propagation time scale of CR nuclei. 
Precise measurements of this isotope would, thus, allow to disentangle the well-known degeneracy between the 
normalization of the diffusion coefficient and the height of the diffusive halo. 
Current measurements of this isotope are still not very precise and bound to very low energies. 
Nonetheless, in the absence of precise $^{10}\mathrm{Be}$ measurements also the total Be flux might allow to draw some conclusions on the height of the diffusive halo.

Deuterium and Helium 3 are also secondaries sensitive to propagation, with  $^3\mathrm{He}$ also recently measured by AMS-02 \cite{Aguilar:2019eiz}. 
Further heavier secondaries are important, in particular sub-Iron species like Scandium (Sc), Titanium (Ti), Vanadium (V), 
Chromium (Cr) and Manganese (Mn), whose measurements from AMS-02 are not yet available.

The aim of the present analysis is to combine the information of multiple primary CRs  C, N and O and 
secondary CRs Li, Be and B in order to derive constraints on the diffusion properties of CR nuclei and 
systematically explore the role of the additional components which might impact in the process of CR propagation. 
We will consider two main complementary propagation frameworks. In the first, CR diffusion is described 
by a smoothly broken power law in rigidity with a break at about $\sim$ 5--10~GeV and no reacceleration. 
The second one, instead, includes reacceleration but it does not  assume a break in diffusion, which is instead replaced by a break in the injection spectrum of primaries.  
Beside these two scenarios, we consider three further frameworks, making it a total of five, that extend the previous analyses in various ways. 
In the most general case we explore a scenario with both reacceleration, a break in diffusion, and a break in injection.

The other ingredient which we will investigate thoroughly is the systematic uncertainty in 
the nuclear cross-section production of secondaries.
Most of these cross sections are poorly measured and thus the effect of their uncertainties on the prediction of the abundance of secondary 
elements is significant. 
This aspect has been stressed in several recent works on CR nuclei
\cite{Genolini:2018ekk,Weinrich:2020cmw,Weinrich:2020ftb,Evoli:2019wwu,Evoli:2019iih,Schroer:2021ojh,Boschini:2018baj,Boschini:2019gow}
as well as for CR antiprotons \cite{Korsmeier:2018gcy,Donato:2017ywo,Winkler:2017xor,Kachelriess:2015wpa}. 
Recently, there has been some experimental effort to gather new measurements of these cross sections \cite{Aaij:2018svt,Unger:2019nus}. 
Furthermore, there is some effort to measure nuclear cross sections directly with the AMS-02 detector itself, see \cite{Aguilar:2021tos}. 
As we will show, taking into account these uncertainties is crucial. In particular, propagation scenarios 
which seem excluded at first remain viable once these uncertainties are considered.
The important element of novelty which we will include in the following will be the simultaneous exploration of the propagation and
cross-section uncertainties in a global fit, with the latter uncertainties included via a parametrization of the cross sections themselves.
The exploration of the large joint parameter space given by propagation and cross-section parameters is a delicate task which we perform
through the use of Monte Carlo scanning techniques.
Furthermore, we check the impact of possible correlations in the AMS-02 data on our results.

The manuscript is structured in the following way: In Sec.~\ref{sec::CR_prop} we give a more general introduction to CR propagation of nuclei in our Galaxy. Then, in Sec.~\ref{sec::data}
we summarize the CR data used in this analysis. Section~\ref{sec::xsc} explains in more detail the importance of fragmentation cross sections for this analysis. 
The methods of the global fits using Monte Carlo scanning techniques are detailed in Sec.~\ref{sec::methods}. 
In this work we will consider 5 different scenarios for CR propagation, which are specified in Sec.~\ref{sec::frameworks}.
Finally, results are presented in Sec.~\ref{sec::results}. More specific results on the halo half-height $z_\mathrm{h}$ are given in \ref{sec::zh}.
Before concluding in Sec.~\ref{sec::conclusions} we discuss and compare our analysis to other works in Sec.~\ref{sec::comparison}.

\section{Cosmic-ray propagation} \label{sec::CR_prop}

Galactic CRs are well described by diffusion within a cylindrical halo extending a few kpc above and below the Galactic plane. 
In general, we distinguish primary and secondary CRs. While primaries are produced and injected into the diffusion halo by 
astrophysical sources like supernova remnants or pulsars, the latter are produced during the process of CR propagation, 
mostly by fragmentation of heavier CR nuclei on the ISM. 
The ISM mostly consists of Hydrogen ($\sim 90$\%) and Helium ($\sim10$\%) gas located in the Galactic disc. 
The whole process of CR propagation can be modeled by a diffusion equation in phase-space densities~\cite{StrongMoskalenko_CR_rewview_2007}:
\begin{eqnarray}
  \label{eqn::propagationEquation}
  \frac{\partial \psi_i (\bm{x}, p, t)}{\partial t} = 
    q_i(\bm{x}, p) &+&  
    \bm{\nabla} \cdot \left(  D_{xx} \bm{\nabla} \psi_i - \bm{V} \psi_i \right) \nonumber \\ 
     &+&  \frac{\partial}{\partial p} p^2 D_{pp} \frac{\partial}{\partial p} \frac{1}{p^2} \psi_i - 
    \frac{\partial}{\partial p} \left( \frac{\diff p}{\diff t} \psi_i  
    - \frac{p}{3} (\bm{\nabla \cdot V}) \psi_i \right) -
    \frac{1}{\tau_{f,i}} \psi_i - \frac{1}{\tau_{r,i}} \psi_i.
\end{eqnarray}
In addition to diffusion several further processes are incorporated in this equation. 
We will briefly mention and define each process in the following. 

The term $q_i(\bm{x}, p)$ denotes the source term of each CR species $i$. For primaries, we factorize the source term 
into a space- and a rigidity-dependent part. Assuming that our Galaxy is cylindrically symmetric we obtain
\begin{equation}
q_i(\bm{x},p) = q_i(r,z,R) = q_{0,i} \, q _{r,z}(r,z) \, q_{i,R}(R)\,,
\end{equation}
where $r$ and $z$ are cylindrical coordinates with respect to the Galactic center and $R=p/Z$ is the rigidity.
The rigidity-dependent part, $q_{i,R}(R)$, is modeled by a smoothly broken power-law 
with a break positioned at $R_{\mathrm{inj},0}$. The spectral indices above and below the break are denoted $\gamma_{1,i}$ and $\gamma_{2,i}$, respectively:
\begin{eqnarray}
  \label{eqn::SourceTerm_2}
  q_R(R) &\sim& R^{-\gamma_1}
                \left( 1+\left( \frac{R}{R_{\mathrm{inj},0}} \right)^{1/s}
                                               \right)^{-s (\gamma_2-\gamma_1)}.
\end{eqnarray}
The parameter $s$ allows for a smoothing of the power law around the break: the larger $s$ the smoother is the transition 
from $\gamma_{1,i}$ to $\gamma_{2,i}$. For completeness, we state also the adopted spatial form of the source term, 
which is the same for all primaries:
\begin{eqnarray}
  \label{eqn::SourceTerm_distribution}
  q_{r,z}(r,z)  = \left( \frac{r}{r_s} \right)^\alpha \exp \left( -\beta \, \frac{r-r_s}{r_s} \right) 
                                                    \exp \left( - \frac{|z|}{z_0} \right),
\end{eqnarray}
with parameters $\alpha = 0.5$, $\beta=2.2$, $r_s=8.5$~kpc, and $z_0=0.2$ kpc.\footnote{These are the default values in \texttt{Galprop v56}, 
which slightly differ from the values obtained from supernova remnants \cite{Case:1998qg,Green:2015isa}. 
However, it was already pointed out in~\cite{Korsmeier:2016kha} that the spatial dependence of 
the source term distribution has only a minor impact on the local CR fluxes.} \\
On the other hand, the source term of secondary CRs is given by: 
\begin{eqnarray}
  \label{eqn::secondary_source_term}
  q_{i}(R_i) &=& \sum\limits_{k>i}\;\sum\limits_{j= \lbrace \mathrm{p,He}\rbrace } 
      4\pi \, n_{\mathrm{ISM},j} \,
      \phi_k(R_k) 
      \,  \sigma_{k+j\rightarrow i}(T_k)\; _{ \Big\rvert_{ \frac{T_k}{A_k}=\frac{T_i}{A_i}} }
\end{eqnarray}
Here $\sigma_{k+j\rightarrow i}$ denotes the fragmentation cross section of species $k$ into $i$ on the ISM component $j$. 
Furthermore, $\phi_k$ is the primary CR flux, and $n_{\mathrm{ISM},j}$ is the density of the ISM gas component $j$. 
The rigidities of the secondary CR $i$ and primary CR $k$ are related by the assumption that the 
kinetic-energy-per nucleon ($T/A$) is conserved during the fragmentation, which is a good approximation~\cite{1995ApJ...451..275T}.
In the above expression for simplicity the spatial dependence of the various quantities has been suppressed but it is taken into account in the calculations.

\bigskip

At the heart of Eq.~\eqref{eqn::propagationEquation} we find the spatial diffusion term. 
The diffusion coefficient $D_{xx}$ is modeled by a smoothly broken power law as function of rigidity $R$ with up to two breaks: 
\begin{eqnarray}
  \label{eqn::diffusionConstant}
  D_{xx} &  \sim& \beta	R^{\delta_l}
  	      \cdot     \left( 1 + \left(\frac{R}{R_{D,0}}\right)^{1/s_{D,0}} \right)^{s_{D,0}\,( \delta - \delta_l) }  
  	      \cdot     \left( 1 + \left(\frac{R}{R_{D,1}}\right)^{1/s_{D,1}} \right)^{s_{D,1}\,( \delta_h - \delta) } . 
\end{eqnarray}
Here $\delta_l$, $\delta$ , $\delta_h$ are the three power law indices below the first break at $R_{D,0}$, between the two breaks, and above the second break at $R_{D,1}$. 
Furthermore, $s_{D,0}$ and $s_{D,1}$ describe the amount of smoothing of the two breaks, respectively, and $\beta=v/c$ is the velocity of the CRs.
The diffusion coefficient is normalized to $D_0$ at a reference rigidity of 4~GV, $D_{xx}( R=4\;\mathrm{GV}) = D_0$. 
The first break, if included in the model, is typically in the range of 1--10 GV. The existence of a second break at higher rigidities 
is suggested by the recent observations of secondary lithium, beryllium, and boron~\cite{Aguilar:2018njt}. 
We will thus include a break around $R_{D,1} \sim 300$~GV with a negligible amount of smoothing, $s_{D,1} \rightarrow 0$. 

Next to diffusion,  reacceleration, convection and energy losses play an important role.
The amount of diffusive reacceleration is parametrized in terms of the velocity $v_{\rm A}$ of Alfv\`en magnetic waves as \cite{Ginzburg:1990sk,1994ApJ...431..705S}:
\begin{eqnarray}
  \label{eqn::DiffusivReaccelerationConstant}
  D_{pp} = \frac{4 \left(p \, v_\mathrm{A} \right)^2 }{3(2-\delta)(2+\delta)(4-\delta)\, \delta \, D_{xx}}.
\end{eqnarray}
The parameter $\delta$ is the power-law index between the breaks from Eq.~\eqref{eqn::diffusionConstant}. 
We assume that the velocity of convective winds, $\bm{V}(\bm{x})$, is constant and orthogonal to the Galactic plane, 
$\bm{V}(\bm{x})= {\rm sign}(z)\, v_{0,{\rm c}}\,{\bm e}_z$. 
Finally, we consider energy losses. These might be continuous, adiabatic or catastrophic. 
Continuous energy losses are encoded in the term $\diff p/\diff t$ and include processes like ionization 
and Coulomb losses, while adiabatic energy losses originate from a nonzero gradient of convective winds.  
Catastrophic energy losses are encoded in the lifetime for fragmentation or decay as described by the parameters  $\tau_f$ and $\tau_r$. 
We note that the catastrophic loss terms give rise to the source term of secondaries (see Eq.~\eqref{eqn::secondary_source_term}).
\medskip

All in all, Eq.~\eqref{eqn::propagationEquation} provides a chain of coupled differential equations, 
where the flux of heavier nuclei (partly) define the source term of lighter secondary CR species. 
It can be solved by starting with heavier species gradually evolving to lighter nuclei. 
We use the \texttt{Galprop} code\footnote{http://galprop.stanford.edu/} \cite{Strong:1998fr,Strong:2015zva} in order to solve the diffusion equation numerically.
More specifically, we use \texttt{Galprop} version~56.0.2870 combined with \textsc{Galtoollibs}~855\footnote{https://galprop.stanford.edu/download.php} 
as basis for our analysis and implement some custom modifications. 
We employ a grid in  kinetic energy per nucleon and in the two spatial dimensions $r$ and $z$, namely we assume cylindrical 
symmetry.\footnote{We note that the diffusion equation can also be solved analytically under various simplifying assumption \cite{Putze:2010zn,Maurin:2018rmm}. 
Moreover, there are  further numerically codes like \texttt{Dragon}~\cite{Evoli:2008dv,Evoli:2017vim} and \texttt{Picard}~\cite{Kissmann:2014sia}.} 
It is assumed that CRs are in a steady state. Our diffusion halo has a maximal extension of $r=20$~kpc and $z=\pm z_\text{h}$. 
Given the well-known degeneracy between $z_\text{h}$ and $D_0$ we will keep $z_\text{h}$ fixed to a benchmark value of 4\;kpc for most of the analysis. 
In a specific paragraph we will vary also  $z_\text{h}$ and discuss its constraints.

\bigskip

Beside propagation in the Galaxy, CRs also propagate in the Solar system. This affects CR spectra only at relatively low rigidities, 
below $\sim 50$ GV, through the effect of solar winds and the solar magnetic field.
This is commonly referred to as solar modulation and varies in strength in a 22-year cycle.
A propagation equation similar to the one of interstellar CR propagation, but adjusted to the  heliospheric conditions, can be used to describe Solar modulation.
The equation can be solved numerically~\cite{Kappl:2015hxv,Maccione:2012cu,Vittino:2017fuh,Boschini:2017gic}. 
There are also approaches to solve approximated versions with semi-anlytical methods\cite{Kuhlen:2019hqb}. 
Progress in the understanding of CR propagation in the heliosphere has been made ~\cite{Tomassetti:2017hbe,Cholis:2015gna} 
particularly thanks to the data of the Voyager~I probe which left the heliosphere a few years ago and 
is thus sampling CRs fluxes before they are affected by the solar modulation.
Beside this, time-dependent measurements of CR fluxes both by PAMELA and 
AMS-02~\cite{Aguilar:2018wmi,Aguilar:2018ons} are also very useful to constrain the properties of heliospheric propagation.
Nonetheless, a detailed understanding is still missing at the moment. 
For this reason we resort to the commonly employed force-field approximation~\cite{Fisk:1976aw}, 
where the CR flux near Earth is given by
\begin{eqnarray}
	\label{eqn::solarModulation}
	\phi_{\oplus,i}(E_{\oplus,i}) &=& \frac{E_{\oplus,i}^2 - m_i^2}{E_{\text{LIS}, i}^2 - m_i^2} \phi_{\text{LIS}, i}(E_{\text{LIS}, i}) \,,\\
	E_{\oplus,i} &=& E_{\text{LIS}, i} - e|Z_i|\varphi_{{\rm SM},i}\,,
\end{eqnarray} 
where $e$ 
is the elementary charge, and $Z_i$,  $m_i$, and $\varphi_{{\rm SM},i}$ are the atomic number, mass, and solar modulation potential of species $i$, respectively. 
$E_{\text{LIS,i}}$ is the CR energy of the local interstellar spectrum, while 
$E_{\oplus,i}$ is the energy effectively measured at the top of the Earth's atmosphere after propagation in the Solar system.
In the force-field approximation $\varphi_{{\rm SM},i}$ is species-independent. Nonetheless, it might be necessary to consider different potentials 
if the measurements for different species are not simultaneous. Fortunately,  the measurements provided by AMS-02 of Li, Be, B, C, N, O which we will use 
in the analysis all refer to the same period of time, thus we will use a single modulation potential for all of the species.

\section{CR Data}\label{sec::data}

In the following we will use the recent AMS-02 measurements of lithium, beryllium and boron \cite{Aguilar:2018njt}, 
carbon and oxygen \cite{Aguilar:2017hno}, and nitrogen \cite{Aguilar:2018keu}. 
All the data refer to the same period of time, namely the first five years of operation. 
This, as mentioned above, allows us to simplify the treatment of solar modulation and assume a single modulation potential for all the species.

For Li, Be and B absolute, fluxes are available. Nonetheless, we will perform fits on the ratios B/C, Li/C and Be/B since in the ratio 
some measurement systematics cancel out and the overall error is smaller. 
Furthermore, while the absolute secondary spectrum depends on the spectrum of the primary and thus on the injection parameters, 
the secondary over primary ratios are very weakly dependent on them, and this helps to speed-up the convergence of the fit.
In Table \ref{tab::data_sets} we report a list of the employed data. 
In the following also VOYAGER data on carbon, oxygen and boron-over-carbon ratio~\cite{2017arXiv171202818W} are shown, 
but it is only used for plotting purposes and not actually included in the analysis.

\begin{table*}
    \caption{CR data sets used in the fits. For each data set we state the experiment, the number of data points, and the reference.
    }
  \centering
\renewcommand{\arraystretch}{1.5}
\begin{tabular}{c @{\hspace{25px}} c @{\hspace{25px}} c @{\hspace{25px}} c @{\hspace{25px}}c @{\hspace{25px}} c}
\hline \hline
{CR species} & {experiment} & {number of data points} &  {Ref.} \\ \hline
C      & AMS-02  &       68       &    \cite{Aguilar:2017hno}       \\
N      & AMS-02  &       66       &    \cite{Aguilar:2018keu}       \\
O      & AMS-02  &       67       &    \cite{Aguilar:2017hno}       \\
B/C    & AMS-02  &       67       &    \cite{Aguilar:2018njt}       \\
Li/C   & AMS-02  &       67       &    \cite{Aguilar:2018njt}       \\
Be/B   & AMS-02  &       66       &    \cite{Aguilar:2018njt}       \\  \hline \hline
\end{tabular}
\renewcommand{\arraystretch}{1.0}
  \label{tab::data_sets}
\end{table*}

\begin{table*}[t!]
    \caption{Summary of cross section related nuisance parameters included in the CR fits. The first column contains 
    the name of the effective fit parameter name. The second column lists the physical parameters related to the effective one and effectively varied in the fit 
    (notation corresponding to Eq.~\ref{eqn::nuisance_XS}). The third 
    column states the sampling procedure, \ie, whether the parameter is sampled by \texttt{MultiNest} or on-the-fly. 
    The last column reports the range of values scanned in the fit.
    Each nuisance parameter is only included if the product species
    is included in the CR fit. We assume that the fragmentation cross sections 
    $X+\mathrm{H}\rightarrow Y$ and $X+\mathrm{He}\rightarrow Y$ are proportional to each other. 
    Thus, we omit the `$+\mathrm{H}$' and `$+\mathrm{He}$' from the notation.
    }
  \centering
\renewcommand{\arraystretch}{1.5}
\begin{tabular}{c @{\hspace{15px}} c @{\hspace{15px}} c @{\hspace{15px}} c}
\hline\hline
{fit parameter} & {nuisance parameters} & {sampling} & {prior} \\ \hline

$\delta_{\mathrm{XS}\rightarrow \mathrm{B}}$ & 
$\delta_{\,^{16}_{\phantom{1}8}\mathrm{O } \rightarrow \,^{10}_{\phantom{1}5}\mathrm{B} } $ \hspace{0.3cm}
$\delta_{\,^{12}_{\phantom{1}6}\mathrm{C } \rightarrow \,^{10}_{\phantom{1}5}\mathrm{B} } $ \hspace{0.3cm}
$\delta_{\,^{16}_{\phantom{1}8}\mathrm{O } \rightarrow \,^{11}_{\phantom{1}5}\mathrm{B} } $ \hspace{0.3cm}
$\delta_{\,^{12}_{\phantom{1}6}\mathrm{C } \rightarrow \,^{11}_{\phantom{1}5}\mathrm{B} } $ &
\texttt{MultiNest} & [-0.3, 0.3]  \\

$\delta_{\mathrm{XS}\rightarrow \mathrm{Li}}$ & 
$\delta_{\,^{16}_{\phantom{1}8}\mathrm{O } \rightarrow \,^{6}_{3}\mathrm{Li} } $ \hspace{0.3cm}
$\delta_{\,^{12}_{\phantom{1}6}\mathrm{C } \rightarrow \,^{6}_{3}\mathrm{Li} } $ \hspace{0.3cm}
$\delta_{\,^{16}_{\phantom{1}8}\mathrm{O } \rightarrow \,^{7}_{3}\mathrm{Li} } $ \hspace{0.3cm}
$\delta_{\,^{12}_{\phantom{1}6}\mathrm{C } \rightarrow \,^{7}_{3}\mathrm{Li} } $  &
\texttt{MultiNest} & [-0.3, 0.3]  \\

$\delta_{\mathrm{XS}\rightarrow \mathrm{Be}}$ & 
$\delta_{\,^{16}_{\phantom{1}8}\mathrm{O } \rightarrow \,^{7}_{4}\mathrm{Be} } $ \hspace{0.3cm}
$\delta_{\,^{12}_{\phantom{1}6}\mathrm{C } \rightarrow \,^{7}_{4}\mathrm{Be} } $ \hspace{0.3cm}
$\delta_{\,^{16}_{\phantom{1}8}\mathrm{O } \rightarrow \,^{9}_{4}\mathrm{Be} } $ \hspace{0.3cm}
$\delta_{\,^{12}_{\phantom{1}6}\mathrm{C } \rightarrow \,^{9}_{4}\mathrm{Be} } $  &
\texttt{MultiNest}  & [-0.3, 0.3] \\

$\delta_{\mathrm{XS}\rightarrow \mathrm{C}}$ & 
$\delta_{\,^{16}_{\phantom{1}8}\mathrm{O } \rightarrow \,^{12}_{\phantom{1}6}\mathrm{C} } $ \hspace{0.3cm}
$\delta_{\,^{16}_{\phantom{1}8}\mathrm{O } \rightarrow \,^{13}_{\phantom{1}6}\mathrm{C} } $  &
\texttt{MultiNest} & [-0.3, 0.3]  \\

$\delta_{\mathrm{XS}\rightarrow \mathrm{N}}$ & 
$\delta_{\,^{16}_{\phantom{1}8}\mathrm{O } \rightarrow \,^{14}_{\phantom{1}7}\mathrm{N} } $ \hspace{0.3cm}
$\delta_{\,^{16}_{\phantom{1}8}\mathrm{O } \rightarrow \,^{15}_{\phantom{1}7}\mathrm{N} } $  &
\texttt{MultiNest} & [-0.3, 0.3]  \\

$A_{\mathrm{XS}\rightarrow \mathrm{B}}$ & 
$A_{\,^{16}_{\phantom{1}8}\mathrm{O } \rightarrow \,^{10}_{\phantom{1}5}\mathrm{B} } $ \hspace{0.3cm}
$A_{\,^{12}_{\phantom{1}6}\mathrm{C } \rightarrow \,^{10}_{\phantom{1}5}\mathrm{B} } $ \hspace{0.3cm}
$A_{\,^{16}_{\phantom{1}8}\mathrm{O } \rightarrow \,^{11}_{\phantom{1}5}\mathrm{B} } $ \hspace{0.3cm}
$A_{\,^{12}_{\phantom{1}6}\mathrm{C } \rightarrow \,^{11}_{\phantom{1}5}\mathrm{B} } $ &
on-the-fly & [0.1, 10]  \\

$A_{\mathrm{XS}\rightarrow \mathrm{Li}}$ & 
$A_{\,^{16}_{\phantom{1}8}\mathrm{O } \rightarrow \,^{6}_{3}\mathrm{Li} } $ \hspace{0.3cm}
$A_{\,^{12}_{\phantom{1}6}\mathrm{C } \rightarrow \,^{6}_{3}\mathrm{Li} } $ \hspace{0.3cm}
$A_{\,^{16}_{\phantom{1}8}\mathrm{O } \rightarrow \,^{7}_{3}\mathrm{Li} } $ \hspace{0.3cm}
$A_{\,^{12}_{\phantom{1}6}\mathrm{C } \rightarrow \,^{7}_{3}\mathrm{Li} } $  &
on-the-fly & [0.1, 10]  \\

$A_{\mathrm{XS}\rightarrow \mathrm{Be}}$ & 
$A_{\,^{16}_{\phantom{1}8}\mathrm{O } \rightarrow \,^{7}_{4}\mathrm{Be} } $ \hspace{0.3cm}
$A_{\,^{12}_{\phantom{1}6}\mathrm{C } \rightarrow \,^{7}_{4}\mathrm{Be} } $ \hspace{0.3cm}
$A_{\,^{16}_{\phantom{1}8}\mathrm{O } \rightarrow \,^{9}_{4}\mathrm{Be} } $ \hspace{0.3cm}
$A_{\,^{12}_{\phantom{1}6}\mathrm{C } \rightarrow \,^{9}_{4}\mathrm{Be} } $ &
on-the-fly & [0.1, 10]   \\

$A_{\mathrm{XS}\rightarrow \mathrm{C}}$ & 
$A_{\,^{16}_{\phantom{1}8}\mathrm{O } \rightarrow \,^{12}_{\phantom{1}6}\mathrm{C} } $ \hspace{0.3cm}
$A_{\,^{16}_{\phantom{1}8}\mathrm{O } \rightarrow \,^{13}_{\phantom{1}6}\mathrm{C} } $  &
\texttt{MultiNest} & [0.5, 1.5]   \\  \vspace{0.3em}

$A_{\mathrm{XS}\rightarrow \mathrm{N}}$ & 
$A_{\,^{16}_{\phantom{1}8}\mathrm{O } \rightarrow \,^{14}_{\phantom{1}7}\mathrm{N} } $ \hspace{0.3cm}
$A_{\,^{16}_{\phantom{1}8}\mathrm{O } \rightarrow \,^{15}_{\phantom{1}7}\mathrm{N} } $  &
\texttt{MultiNest} & [0.5, 1.5]  \\
\hline\hline
\end{tabular}
\renewcommand{\arraystretch}{1.0}
  \label{tab::nuisance_param}
\end{table*}

\section{Nuclear Cross-section}\label{sec::xsc}

The precision of fragmentation cross sections to produce secondary CRs is in many cases significantly worse compared to the precision of recent CR measurements 
provided by the AMS-02 experiment~\cite{Genolini:2018ekk}. Uncertainties are very often at the level of 20-30\% or even more in the cases of cross sections 
for which data are very scarce.
Thus, we allow some flexibility in the default fragmentation cross sections in order to take into account the related uncertainties. 
More specifically, we allow for a freedom in the overall normalization and in the low-energy slope of  each fragmentation  cross section: 
\begin{eqnarray}
  \label{eqn::nuisance_XS} 
  \sigma_{k+j\rightarrow i}(T_k/A)  = \sigma^{\mathrm{default}}_{k+j\rightarrow i} (T_k/A) \cdot A_{k+j\rightarrow i} \cdot
    \begin{cases} 
  	(T_k/A)^{\delta_{k+j\rightarrow i}}  & T_k/A <  T_\mathrm{ref}/A\\
  	1  & \mathrm{otherwise}
  \end{cases}
  \quad . 
\end{eqnarray}
Here $A_{k+j\rightarrow i}$ is an overall renormalization factor and $\delta_{k+j\rightarrow i}$ adjusts the slope 
of the cross section below a reference kinetic energy-per-nucleon chosen to be $  T_\mathrm{ref}/A = 5 \,\mathrm{GeV/nuc}$.  
This choice for $  T_\mathrm{ref}/A$ is justified by the fact that all the cross-section models predict a break around this energy, 
and a flat behavior above it.
As a default model for the cross section we use the \texttt{Galprop} parametrization.\footnote{This corresponds to the option \texttt{kopt=12} in the galdef file.}

The network of reactions involved in the production of the secondary species is extremely large~\cite{Genolini:2018ekk}. 
In order to keep the analysis feasible we thus allow additional freedom only in the cross section of the main reactions contributing to the secondaries, 
namely the ones involving fragmentation of carbon and oxygen~\cite{Genolini:2018ekk}, 
which are the most abundant primary species. Below we briefly list the reactions considered for each species. We also provide
approximate numbers for the relative isotopic abundances of each species, which result from a typical run of \texttt{Galprop}
using the default network of cross sections.

Lithium has two stable isotopes $^{6}_{3}\mathrm{Li} $ and $^{7}_{3}\mathrm{Li}$, roughly equally abundant in CRs. 
We thus consider the four reactions producing these two isotopes from the fragmentation of $^{16}_{\phantom{1}8}\mathrm{O }$ and $^{12}_{\phantom{1}6}\mathrm{C }$. 
There is one subtlety, in fact one should consider eight fragmentation reactions: four on H and four on He. 
The uncertainty on the fragmentation cross sections on He are even larger than the ones involving H. On the other hand, the abundance of He in the ISM is a
factor of 10 smaller than the one of H. We follow the typical assumption, as implemented in \texttt{Galprop}, that the two reactions on H and He are related by a 
rescaling factor.\footnote{The rescaling factor is typically chosen between $A^{2/3}$ and $A$}
Thus, we only introduce one common nuisance parameter for the spallation on H and He.

Beryllium has 3 isotopes $^{7}_{4}\mathrm{Be}$, $^{9}_{4}\mathrm{Be}$ and $^{10}_{\phantom{1}4}\mathrm{Be}$. 
We notice that $^{7}_{4}\mathrm{Be}$ decays by electron capture and is very short lived in atomic form. 
However, the nuclear form as in CRs is stable, apart from a possible capture of electrons in the ISM. 
The electron capture process is implemented in \texttt{Galprop} and we tested that it has a negligible impact on the Be spectrum.
$^{7}_{4}\mathrm{Be}$ and $^{9}_{4}\mathrm{Be}$ are the most abundant CR isotope making between 30\% and 50\% of total beryllium each. 
The less abundant is $^{10}_{\phantom{1}4}\mathrm{Be}$ composes only about 10\% at low energies and 20\% above 100\;GV. 
Thus, we consider nuisance parameters for the production cross sections only for the two most abundant species, $^{7}_{4}\mathrm{Be}$ and $^{9}_{4}\mathrm{Be}$.

Boron has two stable isotopes $^{10}_{\phantom{1}5}\mathrm{B}$ and $^{11}_{\phantom{1}5}\mathrm{B}$ contributing about 1/3 and 2/3, respectively. 
We consider the four related production channels.

CR C is mainly made of the isotope $^{12}_{\phantom{1}6}\mathrm{C }$ with a subdominant part of up to 10\% of $^{13}_{\phantom{1}6}\mathrm{C }$, and it is mainly primary. 
Nonetheless, it has a small secondary component, at the level of $\sim 20$ \%, coming from fragmentation of oxygen. 
The primary part is almost entirely made of $^{12}_{\phantom{1}6}\mathrm{C }$ since $^{13}_{\phantom{1}6}\mathrm{C }$ is 
produced in star synthesis or primordial nucleosynthesis only at the percent level. The secondary part, though, has roughly equal abundances of the two isotopes. 
We thus consider the two reactions producing the two carbon isotopes from oxygen.

Nitrogen also has two stable isotopes, $^{14}_{\phantom{1}7}\mathrm{N}$ and $^{15}_{\phantom{1}7}\mathrm{N}$ and 
has both a primary and secondary component roughly equal in abundance below 50\;GV. At higher energies $^{14}_{\phantom{1}7}\mathrm{N}$ is dominant. 
Similarly to what happens for C, the primary component is almost entirely made of $^{14}_{\phantom{1}7}\mathrm{N}$ since $^{15}_{\phantom{1}7}\mathrm{N}$ 
is only found in nature at the sub-percent level. The secondary component coming from the O fragmentation though has contributions to both isotopes.
We thus consider the two fragmentation reactions. 

Finally,  O is almost purely the isotope $^{16}_{\phantom{1}8}\mathrm{O}$ and can be considered almost entirely as primary, 
with very small traces coming from the fragmentation of heavier nuclei (at the level of less than percent). 
Thus, we do not consider any freedom in the reactions producing oxygen, although in the network we include all the reactions producing it which involve nuclei up to silicon. 
In this respect we stress that all the reactions listed above are the ones in which we allow freedom in the cross sections, and which give  the main contribution to the secondaries, 
typically at the level of 70-80\%. The remaining reactions in the network (which are up to several hundreds, but give overall a subdominant contribution) 
are  all included, although the related cross sections are fixed to the original model.

Table \ref{tab::nuisance_param} lists the nuclear reactions which we will vary in the fit and the related normalization and slope parameters. 
We note that even though we have reduced the number of reactions to a few crucial ones, the number of related parameters is still too large for the analysis to be feasible. 
We thus employ a further simplification and we force the parameters relative to the same element to be all equal. 
In this way we have in total five normalizations and five slopes, i.e. one for each element which is secondary or has a significant secondary component. 
This simplification is justified by the fact that, at the moment, we only have measurements of the global abundance of each element, 
but not yet separate spectra for each isotope. Varying independently the parameters for each isotope would thus give rise to strong 
degeneracies in the parameter space with no added physical value.  
These parameters will be sampled in the fit simultaneously with the CR propagation parameters. 
Just for three of them, listed in the last column of Tab. \ref{tab::nuisance_param} a simplified treatment is possible. 
In fact, the cross-section normalizations of Li, Be and B are basically equivalent to the overall normalization of the spectra itself. 
Thus, they can be marginalized on-the-fly during the fit. This procedure is explained in more detail in the next section.

\begin{table*}[t!]
    \caption{Summary of free CR parameters in the five different frameworks adopted to describe CR propagation. 
    }
  \centering
\renewcommand{\arraystretch}{1.5}
\begin{tabular}{c @{\hspace{10px}} c @{\hspace{10px}} c @{\hspace{10px}} c @{\hspace{10px}}c @{\hspace{10px}} c @{\hspace{10px}} c}
\hline \hline
                                              & BASE               & BASE+$v_A$             & BASE+inj       & BASE+inj+$v_A$     & BASE+inj+$v_A$$-$diff.brk.     & prior                         \\ \hline
$\gamma_1                                $    & $\gamma_1=\gamma_2$& $\gamma_1=\gamma_2$    & free           &  free              & free                         & [0.0, 2.0]                    \\
$\gamma_2                                $    & free               & free                   & free           &  free              & free                         & [2.1, 2.5]                    \\
$R_{\mathrm{inj},0}                      $    & -                  & -                      & free           &  free              & free                         & [1, 10] GV                    \\
$s                                       $    & -                  & -                      & free           &  free              & free                         & [0.1, 0.7]                    \\
$D_0                                     $    & free               & free                   & free           &  free              & free                         & [1e28, 1e29] cm$^2$s$^{-1}$   \\
$\delta_l                                $    & free               & free                   & free           &  free              & $\delta_l=\delta$            & [-1, 0]                       \\
$\delta                                  $    & free               & free                   & free           &  free              & free                         & [0.2, 0.7]                    \\
$\delta_h                                $    & free               & free                   & free           &  free              & free                         & [0.2, 0.7]                    \\
$R_{D,0}                                 $    & free               & free                   & free           &  free              & -                            & [1, 10] GV                    \\
$R_{D,1}                                 $    & free               & free                   & free           &  free              & free                         & [1e5, 5e5] GV                 \\
$s_{D,0}                                 $    & free               & free                   & free           &  free              & -                            & [0.1, 0.7]                    \\
$v_{0,c}                                 $    & free               & free                   & free           &  free              & free                         & [0, 50] km/s                  \\
$v_{A}                                   $    & -                  & free                   & -              &  free              & free                         & [0, 50] km/s                  \\ 
Iso. Ab. $^{12}_{\phantom{1}6}\mathrm{C }$    & free               & free                   & free           &  free              & free                         & [3300, 4000]                  \\ 
Iso. Ab. $^{14}_{\phantom{1}7}\mathrm{N }$    & free               & free                   & free           &  free              & free                         & [200, 500]                    \\ 
Iso. Ab. $^{16}_{\phantom{1}8}\mathrm{O }$    & free               & free                   & free           &  free              & free                         & [4200, 5000]                  \\ 
 $\varphi_{\text{AMS-02}}$                    & free               & free                   & free           &  free              & free                         & 600$\pm$30 MV                 \\ 
\#par                                         &   13               & 14                     & 16             &  17                & 14                                                           \\
\hline \hline
\end{tabular}
\renewcommand{\arraystretch}{1.0}
  \label{tab::frameworks}
\end{table*}

\section{Methods}\label{sec::methods}

The methods used to perform CR fits in this work are very similar to those introduced 
in \cite{Korsmeier:2016kha} and \cite{Cuoco:2019kuu}. While \cite{Korsmeier:2016kha} and several followup 
applications \cite{Cuoco:2016eej, Cuoco:2017rxb, Cuoco:2017iax, Cuoco:2019kuu, Heisig:2020nse} have focused 
on fitting the light CR nuclei, \ie\ protons, helium and antiprotons, here we apply this methodology 
for the first time to heavier nuclei. In particular, we will focus on the nuclei from lithium to oxygen which 
have been measured precisely by the AMS-02 experiment. We will briefly remind the main ingredients of our method as detailed in \cite{Cuoco:2019kuu} 
and then describe the necessary extensions required for this analysis.

As log-likelihood we use a simple $\chi^2$ summed over the various CR species:
\begin{eqnarray}
  	\label{eqn::likelihood_CR}
	-2\,\log({{\cal L}_{{\rm CR}} }) =  \chi^2_{\rm CR} =  \sum\limits_{s} \sum\limits_{i,j} 	
							\left(\phi^{}_{{ s},i}- \phi^{(\text{m})}_{ s} (R_i)\right) 
							\left(\left(\mathcal{V}^{\left(s\right)}\right)^{\!-1} \right)_{ij}
							\left(\phi^{}_{s,j}- \phi^{(\text{m})}_{s} (R_j)\right)\, ,
\end{eqnarray}
where $\phi^{}_{{ s}, i}$ is the experimentally measured flux of the CR species $s$ 
at the rigidity $R_i$ and $\phi^{(\text{m})}_{s}$ is the corresponding  model prediction. 
The covariance matrix $\mathcal{V}^{\left(s\right)}$ contains the uncertainty of the flux measurement. In the 
our fiducial setup we assume uncorrelated uncertainties, i.e.,
$\mathcal{V}_{ij}^{\left( s\right)} = \delta_{ij} \left[\sigma^{}_{{s},i}\right]^2$,
where $\sigma^{}_{{s},i}$ is the error measurement reported by AMS-02 for species $s$ in the rigidity bin $i$. 
More precisely, we sum in quadrature the statistical and the systematic error. 
However, we will also explore, in a few selected cases, the impact of correlated systematics in the experimental uncertainties. 
To derive the full correlation matrix for these cases we will follow the approach of 
Refs.~\cite{Heisig:2020nse}.\footnote{ 
     In more detail, we take the correlations for the B/C ratio from Ref.~\cite{Heisig:2020nse} and derive the covariance matrices for Li/C and Be/B following the same recipe. 
     The covariance matrices for C, N, and O are derived in analogy to cases of $p$ and He in Ref.~\cite{Heisig:2020nse}.
} 
A similar approach was also used in \cite{Derome:2019jfs}.
We will not summarize the procedure here and refer the reader to Refs.~\cite{Derome:2019jfs,Heisig:2020nse} for further details.

The $\chi^2$ and $\phi^\text{(m)}$ have a dependence on the model parameters that for simplicity we have 
suppressed in the above formula. There are two types of parameters.
The first set pertains to CR propagation and is given by a total of 17 parameters.
They are partly described in Sec.~\ref{sec::CR_prop}, but, for convenience, we list them again below.
Four parameters are used to describe the injection spectrum of primaries, 
i.e., the slopes below and above the rigidity break, $\gamma_{1}$, $\gamma_{2}$,
the rigidity break $R_0$ and a smoothing parameter $s$. 
Nine more parameters describe propagation, i.e., the normalization $D_0$ and the slopes
$\delta_l$, $\delta$, and $\delta_h$ of the diffusion coefficient, the break positions and their smoothing 
$R_{D,0}$, $R_{D,1}$, and $s_{D,0}$,\footnote{The smoothing at the second break $s_{D,1}$ is assumed equal to zero and not varied in the fit.} 
the velocity of Alfv\`en magnetic waves, $v_A$, and the convection velocity, $v_{0c}$.
Furthermore, there are 3 parameters to determine the isotopic 
abundance of primary C, N, and O, measured relative to the hydrogen abundance, whose value is arbitrarily fixed to $1.06\times10^6$.
The last parameter is the solar modulation potential $\varphi_{\text{AMS-02}}$.
During several preliminary fits we left these parameters completely free to vary and we found that the preferred value is about 600 MV.
For better stability of the subsequent fits, we then decided to
apply a Gaussian prior on $\varphi_{\text{AMS-02}}$, \ie, we add to the main likelihood the term
$-2\,\log({{\cal L}_{{\rm SM}} }) = (\varphi_{\text{AMS-02}}- 600\,\mathrm{MV} )^2 / \sigma_\varphi^2$
where $\sigma_\varphi = 30$ MV.
Not all of these parameters are present in each fit. Depending on the propagation scenario under consideration (see below) only a subset might be present. 
These parameters, together with the prior ranges explored in the fit are reported in Tab.~\ref{tab::frameworks}.

The second set of parameters is related to the nuclear fragmentation cross sections. 
As described in Sec.~\ref{sec::xsc} we consider a total of 10 cross-section parameters. 
Also in this case, not all of the parameters are necessarily present in each fit. 
A fit with BCNO for example only will use 6 parameters, while the fit with all of the species will include all 10 parameters. 
As mentioned in Sec.~\ref{sec::xsc} the 3 normalization parameters of Li, Be and B 
are treated as on-the-fly marginalization parameters, as explained further below.

To scan this large combined propagation-cross-section parameter space (up to 27-dimensional in the largest case) 
we use \texttt{MultiNest}~\cite{Feroz:2008xx}.
For the \texttt{MultiNest} setup we use 400 live points, an enlargement factor \textsc{efr=0.7},
and a stopping criterion of \textsc{tol=0.1}.
As mentioned above some parameters are treated in a simplified way, namely the modulation potential and three cross-section normalizations. 
We profile over these four parameters together on-the-fly at each \texttt{MultiNest} likelihood evaluation following~\cite{Rolke:2004mj}.
More precisely, for each evaluation in the fit within the parameter space scanned by \texttt{MultiNest}
the likelihood is maximized over the four remaining simplified parameters. This maximization is performed
with \texttt{Minuit}~\cite{James:1975dr}.
We remark that the parameter scans performed in this analysis require a large computational effort. A typical \texttt{MultiNest} scan requires 0.5 to 1\;M likelihood evaluations. 
However, depending on the complexity of the fit this number can increase significantly. For example, when the half-height of the diffusion halo is further included as free parameter (\cf\ Sec.~\ref{sec::zh}), 
the fit has to explore the additional $D_0-z_\text{h}$ degeneracy. Thus, in some cases the number of required likelihood evaluations can exceed 3\;M.
A single likelihood evaluation takes between 150 and 200 sec on a single core.

\texttt{MultiNest} formally explores the Bayesian posterior and thus naturally provides sample from it. Nonetheless, these samples can also be used
to perform a frequentist statistical analysis of the results.
As default choice to interpret the scan result we use a frequentist framework, and we build one and two-dimensional
profile likelihoods in the different parameters, from which we derive contours which are shown in the
various figures in the following. Nonetheless, we also check the analysis results given by a Bayesian approach.
We find that two interpretations give almost indistinguishable constraints indicating that the data are constraining enough
to make the results basically  prior-independent, and statical interpretation independent.
An explicit example comparison is given in the Appendix.

\section{Five frameworks to describe CR propagation}\label{sec::frameworks}

In Sec.~\ref{sec::CR_prop} we have described a very general approach to model CR propagation in this work. 
However, one of the key motivations of this analysis is to check whether the new precise AMS-02 measurements of several CR secondary spectra 
can point to a more specific framework of CR propagation. 
Therefore, we define, select and study five propagation scenarios,  four of which are sub-cases of the most general one. The parameters of the various scenarios 
together with the range of variation explored in the fits are listed in Tab.~\ref{tab::frameworks}.

The most simple framework used in this analysis, \ie, the one with the smaller number of parameters, is called BASE. 
In this framework there is no break in the injection spectrum of primary CRs which thus follows a simple power law in rigidity with spectral index  $\gamma_2$. 
On the other hand, the diffusion coefficient is modeled by a double-broken power law. The first break is at a few GVs. 
A break in this range is motivated by various theoretical studies \cite{Ptuskin:2005ax,Blasi:2012yr}. 
While a second break around few hundreds of GVs is favoured by the recent measurements of AMS-02 of Li, Be and B \cite{Aguilar:2018njt}. 
Finally, in this scenario we also allow for convection, while reacceleration is forced to be zero. 
In total this model has 9 free propagation parameters. To them we then need to add C, N, and O abundances, 
the solar modulation potential and the cross-section uncertainty parameters.

The BASE scenario is gradually complicated by adding further ingredients. We consider the BASE+$v_A$ scenario 
which is identical to the BASE one but further includes reacceleration. It thus have one additional parameter. 
The BASE+inj scenario, instead, it goes beyond the BASE scenario including a break in the injection spectrum of primaries. 
It thus has three parameters more, namely $\gamma_1$, $R_{\mathrm{inj},0}$ and $s$. 
The most general model is given by an extension of the BASE framework by both reacceleration and 
a break in the injection spectrum which we dub BASE+inj+$v_A$.It has four more parameters with respect to BASE.
Finally, the last scenario we consider includes reacceleration and a break in the injection spectrum 
but we remove the break in the diffusion coefficient at rigidities of a few GV.
We refer to this framework as BASE+inj+$v_A$$-$diff.brk. This last scenario corresponds to the one explored 
in our previous analyses~\cite{Korsmeier:2016kha, Cuoco:2016eej, Cuoco:2017rxb, Cuoco:2017iax, Cuoco:2019kuu, Heisig:2020nse} 
as well as in other works~\cite{Trotta:2010mx,Johannesson:2016rlh,Boschini:2018baj}.   
Regarding the dataset employed, every fit includes  AMS-02 data on C, N and O. Fits can then be grouped into two categories: 
In the first one we include the B/C ratio dubbing this configuration BCNO, 
while in the second group we include also the Li/C ratio and the Be/B ratio, \ie, all the nuclei considered in this work, dubbing this 
configuration as LiBeBCNO.
All in all, this gives 10 CR fits corresponding to 5 different frameworks of CR propagation and 2 CR data set configurations. 
Finally,  only for the BASE and BASE+inj+$v_A$$-$diff.brk frameworks with BCNO and LiBeBCNO data configurations we consider four more fits,  
to study the effect of correlated systematic uncertainties. In total we will thus perform 14 fits.

\begin{sidewaystable}    
   \caption{
        Fit results. For all 14 fits we report the total $\chi^2$, the contribution to the $\chi^2$ form each single species, 
        the number of degrees of freedom, and the best-fit value and 1 $\sigma$ error for each parameter\@.
    }
  \centering
\scalebox{0.77}{
\renewcommand{\arraystretch}{1.5}
\begin{tabular}{ccccccccccccccc}
\hline \hline
                                              &  \multicolumn{2}{c}{ $\mathrm{BASE}$ }                                                  & \multicolumn{2}{c}{ $\mathrm{BASE (corr)}$ }                                            & \multicolumn{2}{c}{ $\mathrm{BASE}+v_A$ }                                               & \multicolumn{2}{c}{ $\mathrm{BASE}+\mathrm{inj}$ }                                      & \multicolumn{2}{c}{ $\mathrm{BASE}+\mathrm{inj}+v_A$ }                                  &\multicolumn{2}{c}{ $\mathrm{BASE}+\mathrm{inj}+v_A-\mathrm{diff.brk.}$      }           &\multicolumn{2}{c}{ $\mathrm{BASE}+\mathrm{inj}+v_A-\mathrm{diff.brk.}$ (corr) }         \\ \hline
data set                                      & BCNO                                       & LiBeBCNO                                   & BCNO                                       & LiBeBCNO                                   & BCNO                                       & LiBeBCNO                                   & BCNO                                       & LiBeBCNO                                   & BCNO                                       & LiBeBCNO                                   & BCNO                                       & LiBeBCNO                                   & BCNO                                       & LiBeBCNO                                   \\
$\#\mathrm{dof}                              $& $                                     252$ & $                                     383$ & $                                     252$ & $                                     383$ & $                                     251$ & $                                     382$ & $                                     249$ & $                                     380$ & $                                     248$ & $                                     379$ & $                                     251$ & $                                     382$ & $                                     251$ & $                                     382$ \\
$\chi^2                                      $& $                                    72.4$ & $                                   170.0$ & $                                   423.2$ & $                                   593.6$ & $                                    72.8$ & $                                   169.0$ & $                                    67.9$ & $                                   160.7$ & $                                    67.3$ & $                                   158.9$ & $                                    74.2$ & $                                   168.9$ & $                                   415.4$ & $                                   590.2$ \\
$\chi^2_\mathrm{N}                           $& $                                    15.9$ & $                                    18.7$ & $                                   148.1$ & $                                   146.9$ & $                                    19.3$ & $                                    17.2$ & $                                    14.8$ & $                                    15.2$ & $                                    19.3$ & $                                    16.5$ & $                                    17.2$ & $                                    20.0$ & $                                   151.5$ & $                                   149.1$ \\
$\chi^2_\mathrm{O}                           $& $                                    14.0$ & $                                    13.7$ & $                                    61.8$ & $                                    63.0$ & $                                    11.3$ & $                                    14.8$ & $                                    12.8$ & $                                    13.8$ & $                                    11.2$ & $                                    12.4$ & $                                    14.8$ & $                                    12.1$ & $                                    62.1$ & $                                    62.2$ \\
$\chi^2_\mathrm{C}                           $& $                                    13.1$ & $                                    15.2$ & $                                   127.8$ & $                                   124.9$ & $                                    12.7$ & $                                    12.7$ & $                                    16.0$ & $                                    16.9$ & $                                    11.7$ & $                                    15.6$ & $                                    13.7$ & $                                    15.4$ & $                                   122.4$ & $                                   122.5$ \\
$\chi^2_\mathrm{Be/B}                        $& -                                          & $                                    42.2$ & -                                          & $                                    82.9$ & -                                          & $                                    42.0$ & -                                          & $                                    43.3$ & -                                          & $                                    42.4$ & -                                          & $                                    40.6$ & -                                          & $                                    83.3$ \\
$\chi^2_\mathrm{Li/C}                        $& -                                          & $                                    46.5$ & -                                          & $                                    82.5$ & -                                          & $                                    47.5$ & -                                          & $                                    39.1$ & -                                          & $                                    39.2$ & -                                          & $                                    41.2$ & -                                          & $                                    85.3$ \\
$\chi^2_\mathrm{B/C}                         $& $                                    27.8$ & $                                    30.2$ & $                                    78.4$ & $                                    80.5$ & $                                    28.7$ & $                                    29.5$ & $                                    24.1$ & $                                    26.9$ & $                                    24.0$ & $                                    28.1$ & $                                    25.8$ & $                                    33.4$ & $                                    75.8$ & $                                    82.7$ \\
$\gamma_1                                    $& -                                          & -                                          & -                                          & -                                          & -                                          & -                                          & $                  {2.18}^{+0.04}_{-0.51}$ & $                  {2.21}^{+0.04}_{-0.07}$ & $                  {2.08}^{+0.10}_{-0.30}$ & $                  {2.10}^{+0.10}_{-0.06}$ & $                  {1.20}^{+0.42}_{-0.16}$ & $                  {1.64}^{+0.04}_{-0.07}$ & $                  {1.15}^{+0.08}_{-0.12}$ & $                  {1.14}^{+0.24}_{-0.11}$ \\
$\gamma_2                                    $& $               {2.357}^{+0.003}_{-0.005}$ & $               {2.365}^{+0.005}_{-0.002}$ & $                  {2.34}^{+0.01}_{-0.01}$ & $               {2.360}^{+0.014}_{-0.009}$ & $               {2.353}^{+0.006}_{-0.004}$ & $               {2.361}^{+0.009}_{-0.002}$ & $               {2.368}^{+0.002}_{-0.017}$ & $               {2.371}^{+0.004}_{-0.005}$ & $               {2.360}^{+0.008}_{-0.004}$ & $               {2.378}^{+0.003}_{-0.005}$ & $               {2.362}^{+0.016}_{-0.004}$ & $               {2.389}^{+0.005}_{-0.004}$ & $               {2.365}^{+0.008}_{-0.020}$ & $               {2.373}^{+0.013}_{-0.005}$ \\
$R_{inj,0} \;\mathrm{[10^{3}\;     MV]}      $& -                                          & -                                          & -                                          & -                                          & -                                          & -                                          & $    {8.85}^{+1.15}_{-4.33}              $ & $    {8.31}^{+0.91}_{-1.05}              $ & $    {6.20}^{+1.87}_{-1.84}              $ & $    {6.98}^{+2.10}_{-0.33}              $ & $    {3.28}^{+1.82}_{-0.59}              $ & $    {5.18}^{+0.65}_{-0.30}              $ & $    {2.93}^{+0.20}_{-0.37}              $ & $    {2.61}^{+1.01}_{-0.10}              $ \\
$s                                           $& -                                          & -                                          & -                                          & -                                          & -                                          & -                                          & $                  {0.48}^{+0.02}_{-0.27}$ & $                  {0.45}^{+0.04}_{-0.06}$ & $                  {0.45}^{+0.03}_{-0.20}$ & $               {0.487}^{+0.006}_{-0.043}$ & $               {0.490}^{+0.009}_{-0.052}$ & $               {0.493}^{+0.007}_{-0.044}$ & $                  {0.39}^{+0.10}_{-0.06}$ & $               {0.494}^{+0.005}_{-0.038}$ \\
$D_0\;\mathrm{[10^{28}\;      cm^2/s]}       $& $   {5.05}^{+0.99}_{-1.34}               $ & $   {4.24}^{+0.96}_{-0.44}               $ & $   {4.08}^{+0.33}_{-0.55}               $ & $   {4.01}^{+0.32}_{-0.45}               $ & $   {4.62}^{+1.28}_{-0.46}               $ & $   {4.52}^{+1.94}_{-0.39}               $ & $   {3.82}^{+1.03}_{-0.30}               $ & $   {3.73}^{+0.69}_{-0.20}               $ & $   {3.65}^{+0.87}_{-0.18}               $ & $   {3.53}^{+0.25}_{-0.10}               $ & $   {4.16}^{+0.33}_{-0.88}               $ & $   {4.34}^{+0.14}_{-0.66}               $ & $   {3.24}^{+0.94}_{-0.36}               $ & $   {3.60}^{+0.19}_{-0.33}               $ \\
$\delta_l                                    $& $                 {-0.98}^{+0.22}_{-0.01}$ & $              {-0.997}^{+0.111}_{-0.001}$ & $                 {-0.98}^{+0.03}_{-0.02}$ & $                 {-0.97}^{+0.03}_{-0.02}$ & $                 {-0.91}^{+0.11}_{-0.07}$ & $                 {-0.91}^{+0.08}_{-0.09}$ & $                 {-0.57}^{+0.22}_{-0.37}$ & $                 {-0.70}^{+0.09}_{-0.28}$ & $                 {-0.88}^{+0.36}_{-0.06}$ & $                 {-0.91}^{+0.13}_{-0.04}$ & -                                          & -                                          & -                                          & -                                          \\
$\delta                                      $& $                  {0.49}^{+0.03}_{-0.04}$ & $               {0.499}^{+0.002}_{-0.033}$ & $                  {0.48}^{+0.03}_{-0.01}$ & $                  {0.47}^{+0.02}_{-0.01}$ & $               {0.498}^{+0.007}_{-0.045}$ & $               {0.496}^{+0.004}_{-0.056}$ & $                  {0.47}^{+0.02}_{-0.03}$ & $                  {0.48}^{+0.01}_{-0.03}$ & $                  {0.47}^{+0.02}_{-0.03}$ & $               {0.471}^{+0.009}_{-0.014}$ & $                  {0.45}^{+0.02}_{-0.02}$ & $               {0.414}^{+0.013}_{-0.005}$ & $                  {0.47}^{+0.02}_{-0.04}$ & $                  {0.43}^{+0.02}_{-0.01}$ \\
$\delta_h                                    $& $               {0.315}^{+0.045}_{-0.008}$ & $               {0.340}^{+0.007}_{-0.033}$ & $                  {0.32}^{+0.03}_{-0.02}$ & $               {0.293}^{+0.032}_{-0.009}$ & $                  {0.33}^{+0.02}_{-0.02}$ & $               {0.331}^{+0.008}_{-0.027}$ & $                  {0.31}^{+0.03}_{-0.02}$ & $                  {0.33}^{+0.02}_{-0.03}$ & $                  {0.31}^{+0.03}_{-0.01}$ & $                  {0.31}^{+0.01}_{-0.01}$ & $                  {0.30}^{+0.04}_{-0.02}$ & $               {0.271}^{+0.026}_{-0.007}$ & $                  {0.31}^{+0.02}_{-0.03}$ & $               {0.311}^{+0.007}_{-0.044}$ \\
$R_{D,0}  \;\mathrm{[10^{3}\;     MV]}       $& $    {3.94}^{+0.52}_{-0.35}              $ & $    {4.05}^{+0.43}_{-0.14}              $ & $    {3.87}^{+0.14}_{-0.12}              $ & $    {3.85}^{+0.16}_{-0.05}              $ & $    {3.97}^{+0.21}_{-0.36}              $ & $    {4.25}^{+0.10}_{-0.35}              $ & $    {4.07}^{+0.20}_{-0.53}              $ & $    {4.01}^{+0.14}_{-0.37}              $ & $    {3.02}^{+0.81}_{-0.23}              $ & $    {3.37}^{+0.43}_{-0.41}              $ & -                                          & -                                          & -                                          & -                                          \\
$R_{D,1}  \;\mathrm{[10^{5}\;     MV]}       $& $    {1.80}^{+0.13}_{-0.30}              $ & $    {1.52}^{+0.48}_{-0.08}              $ & $    {2.00}^{+0.25}_{-0.22}              $ & $    {2.09}^{+0.14}_{-0.42}              $ & $    {1.88}^{+0.12}_{-0.35}              $ & $    {1.63}^{+0.19}_{-0.07}              $ & $    {1.65}^{+0.35}_{-0.13}              $ & $    {1.49}^{+0.36}_{-0.06}              $ & $    {2.02}^{+0.09}_{-0.46}              $ & $    {1.68}^{+0.12}_{-0.08}              $ & $    {2.14}^{+0.16}_{-0.40}              $ & $    {2.33}^{+0.16}_{-0.46}              $ & $    {1.96}^{+0.62}_{-0.11}              $ & $    {2.12}^{+0.25}_{-0.29}              $ \\
$s_{D,0}                                     $& $                  {0.38}^{+0.06}_{-0.11}$ & $                  {0.32}^{+0.06}_{-0.07}$ & $                  {0.15}^{+0.03}_{-0.02}$ & $                  {0.16}^{+0.03}_{-0.01}$ & $                  {0.36}^{+0.06}_{-0.07}$ & $                  {0.31}^{+0.13}_{-0.05}$ & $                  {0.12}^{+0.19}_{-0.02}$ & $                  {0.13}^{+0.06}_{-0.02}$ & $                  {0.13}^{+0.20}_{-0.02}$ & $               {0.109}^{+0.033}_{-0.004}$ & -                                          & -                                          & -                                          & -                                          \\
$v_{0,c}\;\mathrm{[km/s]}                    $& $                 {3.34}^{+21.76}_{-2.49}$ & $                 {1.81}^{+17.74}_{-0.70}$ & $                  {9.09}^{+7.89}_{-8.68}$ & $                 {12.11}^{+5.83}_{-6.91}$ & $                 {0.27}^{+23.83}_{-0.06}$ & $                 {0.84}^{+27.41}_{-0.22}$ & $               {13.18}^{+14.33}_{-12.26}$ & $                 {4.92}^{+10.66}_{-4.85}$ & $                 {5.02}^{+18.32}_{-2.27}$ & $                  {2.30}^{+6.45}_{-1.31}$ & $                  {0.34}^{+3.88}_{-0.23}$ & $               {0.004}^{+1.515}_{-0.000}$ & $                  {0.89}^{+5.05}_{-0.75}$ & $                  {1.81}^{+2.30}_{-1.63}$ \\
$v_{A}\;\mathrm{[km/s]}                      $& -                                          & -                                          & -                                          & -                                          & $                  {8.65}^{+3.51}_{-7.81}$ & $                  {0.54}^{+6.04}_{-0.24}$ & -                                          & -                                          & $                 {10.68}^{+2.94}_{-9.29}$ & $                 {10.85}^{+3.55}_{-4.79}$ & $                 {19.23}^{+3.65}_{-3.77}$ & $                 {24.04}^{+0.91}_{-2.90}$ & $                 {16.24}^{+5.30}_{-1.35}$ & $                 {20.14}^{+1.44}_{-1.49}$ \\
$\mathrm{Iso. Ab.\,C \;[10^{3}]}             $& $    {3.59}^{+0.08}_{-0.02}              $ & $    {3.59}^{+0.04}_{-0.02}              $ & $    {3.48}^{+0.03}_{-0.14}              $ & $    {3.37}^{+0.11}_{-0.06}              $ & $    {3.63}^{+0.02}_{-0.04}              $ & $    {3.60}^{+0.03}_{-0.02}              $ & $    {3.58}^{+0.05}_{-0.03}              $ & $    {3.59}^{+0.03}_{-0.04}              $ & $ {3.640}^{+0.009}_{-0.068}              $ & $    {3.57}^{+0.03}_{-0.02}              $ & $    {3.58}^{+0.06}_{-0.04}              $ & $    {3.54}^{+0.05}_{-0.01}              $ & $    {3.47}^{+0.08}_{-0.12}              $ & $    {3.36}^{+0.16}_{-0.02}              $ \\
$\mathrm{Iso. Ab.\,N}                        $& $               {325.38}^{+17.75}_{-6.27}$ & $               {306.87}^{+17.12}_{-7.38}$ & $              {276.35}^{+44.56}_{-20.91}$ & $              {280.12}^{+23.61}_{-35.03}$ & $               {348.86}^{+7.40}_{-25.27}$ & $               {323.27}^{+9.25}_{-17.15}$ & $              {333.27}^{+23.85}_{-21.66}$ & $               {307.74}^{+18.80}_{-8.56}$ & $               {327.91}^{+14.14}_{-8.92}$ & $               {313.82}^{+8.68}_{-16.42}$ & $              {337.18}^{+26.21}_{-38.86}$ & $              {300.77}^{+14.11}_{-14.55}$ & $              {308.24}^{+16.04}_{-49.55}$ & $               {228.85}^{+61.43}_{-7.77}$ \\
$\mathrm{Iso. Ab.\,O \;[10^{3}]}             $& $    {4.35}^{+0.18}_{-0.02}              $ & $    {4.41}^{+0.05}_{-0.04}              $ & $    {4.40}^{+0.05}_{-0.10}              $ & $    {4.40}^{+0.05}_{-0.08}              $ & $    {4.41}^{+0.05}_{-0.05}              $ & $    {4.41}^{+0.03}_{-0.09}              $ & $    {4.38}^{+0.05}_{-0.04}              $ & $    {4.37}^{+0.04}_{-0.07}              $ & $    {4.42}^{+0.03}_{-0.09}              $ & $    {4.34}^{+0.07}_{-0.01}              $ & $ {4.313}^{+0.181}_{-0.004}              $ & $    {4.34}^{+0.11}_{-0.02}              $ & $    {4.32}^{+0.23}_{-0.01}              $ & $    {4.41}^{+0.12}_{-0.05}              $ \\
$\delta_\mathrm{XS} \rightarrow \mathrm{C}   $& $                 {-0.08}^{+0.23}_{-0.08}$ & $                  {0.03}^{+0.14}_{-0.13}$ & $                  {0.17}^{+0.09}_{-0.13}$ & $                  {0.13}^{+0.14}_{-0.07}$ & $                 {-0.05}^{+0.08}_{-0.05}$ & $                 {-0.12}^{+0.18}_{-0.06}$ & $                  {0.17}^{+0.06}_{-0.21}$ & $                  {0.11}^{+0.12}_{-0.07}$ & $                  {0.23}^{+0.03}_{-0.24}$ & $                  {0.15}^{+0.05}_{-0.07}$ & $                  {0.28}^{+0.02}_{-0.09}$ & $                  {0.25}^{+0.04}_{-0.03}$ & $                  {0.26}^{+0.03}_{-0.06}$ & $                  {0.22}^{+0.08}_{-0.09}$ \\
$\delta_\mathrm{XS} \rightarrow \mathrm{N}   $& $                 {-0.08}^{+0.07}_{-0.03}$ & $                 {-0.06}^{+0.04}_{-0.04}$ & $                  {0.15}^{+0.02}_{-0.04}$ & $                  {0.12}^{+0.03}_{-0.02}$ & $                 {-0.10}^{+0.06}_{-0.01}$ & $                 {-0.06}^{+0.03}_{-0.05}$ & $                  {0.02}^{+0.06}_{-0.06}$ & $                  {0.01}^{+0.02}_{-0.02}$ & $                  {0.05}^{+0.02}_{-0.07}$ & $               {0.050}^{+0.009}_{-0.032}$ & $                  {0.10}^{+0.02}_{-0.04}$ & $               {0.110}^{+0.005}_{-0.034}$ & $               {0.189}^{+0.008}_{-0.037}$ & $               {0.189}^{+0.004}_{-0.045}$ \\
$\delta_\mathrm{XS} \rightarrow \mathrm{Li}  $& -                                          & $                  {0.00}^{+0.05}_{-0.03}$ & -                                          & $                  {0.16}^{+0.03}_{-0.03}$ & -                                          & $                 {-0.02}^{+0.05}_{-0.06}$ & -                                          & $                  {0.14}^{+0.03}_{-0.02}$ & -                                          & $                  {0.16}^{+0.01}_{-0.02}$ & -                                          & $               {0.193}^{+0.007}_{-0.005}$ & -                                          & $               {0.190}^{+0.005}_{-0.014}$ \\
$\delta_\mathrm{XS} \rightarrow \mathrm{Be}  $& -                                          & $                  {0.07}^{+0.05}_{-0.04}$ & -                                          & $               {0.186}^{+0.010}_{-0.044}$ & -                                          & $                  {0.07}^{+0.04}_{-0.05}$ & -                                          & $                  {0.22}^{+0.02}_{-0.04}$ & -                                          & $                  {0.23}^{+0.02}_{-0.02}$ & -                                          & $               {0.280}^{+0.008}_{-0.012}$ & -                                          & $                  {0.27}^{+0.01}_{-0.03}$ \\
$\delta_\mathrm{XS} \rightarrow \mathrm{B}   $& $              {-0.065}^{+0.084}_{-0.008}$ & $                 {-0.06}^{+0.03}_{-0.02}$ & $                  {0.07}^{+0.02}_{-0.03}$ & $               {0.066}^{+0.009}_{-0.025}$ & $                 {-0.05}^{+0.03}_{-0.02}$ & $                 {-0.07}^{+0.03}_{-0.03}$ & $                  {0.05}^{+0.05}_{-0.05}$ & $                  {0.04}^{+0.01}_{-0.02}$ & $                  {0.09}^{+0.01}_{-0.07}$ & $                  {0.05}^{+0.02}_{-0.01}$ & $                  {0.16}^{+0.03}_{-0.04}$ & $               {0.117}^{+0.005}_{-0.014}$ & $                  {0.12}^{+0.03}_{-0.02}$ & $                  {0.12}^{+0.02}_{-0.02}$ \\
$A_\mathrm{XS}      \rightarrow \mathrm{C}   $& $                  {0.55}^{+0.04}_{-0.04}$ & $                  {0.57}^{+0.03}_{-0.06}$ & $                  {0.63}^{+0.37}_{-0.07}$ & $                  {0.81}^{+0.11}_{-0.27}$ & $                  {0.54}^{+0.05}_{-0.03}$ & $                  {0.54}^{+0.05}_{-0.04}$ & $                  {0.51}^{+0.15}_{-0.01}$ & $                  {0.53}^{+0.05}_{-0.02}$ & $               {0.511}^{+0.074}_{-0.006}$ & $               {0.514}^{+0.031}_{-0.006}$ & $                  {0.54}^{+0.04}_{-0.04}$ & $                  {0.52}^{+0.03}_{-0.02}$ & $               {0.500}^{+0.334}_{-0.000}$ & $                  {0.83}^{+0.06}_{-0.28}$ \\
$A_\mathrm{XS}      \rightarrow \mathrm{N}   $& $                  {1.18}^{+0.04}_{-0.16}$ & $                  {1.11}^{+0.04}_{-0.04}$ & $                  {1.18}^{+0.10}_{-0.13}$ & $                  {1.15}^{+0.10}_{-0.09}$ & $                  {1.09}^{+0.07}_{-0.03}$ & $                  {1.13}^{+0.06}_{-0.04}$ & $                  {1.07}^{+0.13}_{-0.03}$ & $                  {1.15}^{+0.04}_{-0.04}$ & $                  {1.14}^{+0.06}_{-0.04}$ & $                  {1.13}^{+0.04}_{-0.03}$ & $                  {1.19}^{+0.03}_{-0.17}$ & $                  {1.20}^{+0.01}_{-0.13}$ & $                  {1.02}^{+0.17}_{-0.05}$ & $                  {1.15}^{+0.04}_{-0.15}$ \\
$A_\mathrm{XS}      \rightarrow \mathrm{Li}  $& -                                          & $                  {1.27}^{+0.07}_{-0.04}$ & -                                          & $                  {1.22}^{+0.07}_{-0.06}$ & -                                          & $                  {1.32}^{+0.10}_{-0.05}$ & -                                          & $                  {1.33}^{+0.06}_{-0.04}$ & -                                          & $                  {1.31}^{+0.04}_{-0.05}$ & -                                          & $                  {1.36}^{+0.02}_{-0.15}$ & -                                          & $                  {1.14}^{+0.04}_{-0.07}$ \\
$A_\mathrm{XS}      \rightarrow \mathrm{Be}  $& -                                          & $               {0.990}^{+0.006}_{-0.004}$ & -                                          & $               {0.946}^{+0.004}_{-0.003}$ & -                                          & $               {0.992}^{+0.005}_{-0.003}$ & -                                          & $               {0.990}^{+0.005}_{-0.002}$ & -                                          & $               {0.991}^{+0.002}_{-0.003}$ & -                                          & $               {0.992}^{+0.001}_{-0.005}$ & -                                          & $               {0.941}^{+0.007}_{-0.009}$ \\
$A_\mathrm{XS}      \rightarrow \mathrm{B}   $& $                  {1.11}^{+0.04}_{-0.13}$ & $                  {1.03}^{+0.04}_{-0.03}$ & $                  {1.05}^{+0.09}_{-0.08}$ & $                  {1.01}^{+0.05}_{-0.05}$ & $                  {1.08}^{+0.05}_{-0.04}$ & $                  {1.06}^{+0.07}_{-0.03}$ & $                  {1.04}^{+0.11}_{-0.02}$ & $                  {1.06}^{+0.04}_{-0.03}$ & $                  {1.08}^{+0.05}_{-0.03}$ & $                  {1.05}^{+0.03}_{-0.03}$ & $                  {1.16}^{+0.01}_{-0.17}$ & $                  {1.08}^{+0.01}_{-0.10}$ & $                  {0.95}^{+0.11}_{-0.06}$ & $                  {0.95}^{+0.02}_{-0.05}$ \\ \vspace{0.3em} 
$\varphi_\mathrm{AMS-02}                     $& $              {613.67}^{+44.11}_{-13.66}$ & $              {615.56}^{+34.46}_{-16.99}$ & $              {678.15}^{+16.99}_{-21.33}$ & $               {697.68}^{+2.03}_{-21.99}$ & $               {608.98}^{+34.11}_{-7.58}$ & $              {614.39}^{+24.90}_{-17.40}$ & $              {594.81}^{+31.64}_{-11.66}$ & $              {614.82}^{+12.74}_{-40.06}$ & $              {618.80}^{+15.41}_{-25.78}$ & $               {593.84}^{+13.47}_{-8.50}$ & $              {590.95}^{+56.33}_{-13.03}$ & $              {600.02}^{+12.65}_{-23.88}$ & $              {655.17}^{+17.48}_{-45.23}$ & $              {658.61}^{+27.11}_{-36.18}$ \\
\hline
\hline
\end{tabular}
}
\renewcommand{\arraystretch}{1.0}
  \label{tab::fit_results}
\end{sidewaystable}

\begin{figure*}[!b]
\centering
\setlength{\unitlength}{1\textwidth}
\begin{picture}(1,0.75)
 \put(-0.025, -0.0){\includegraphics[width=0.25\textwidth ]{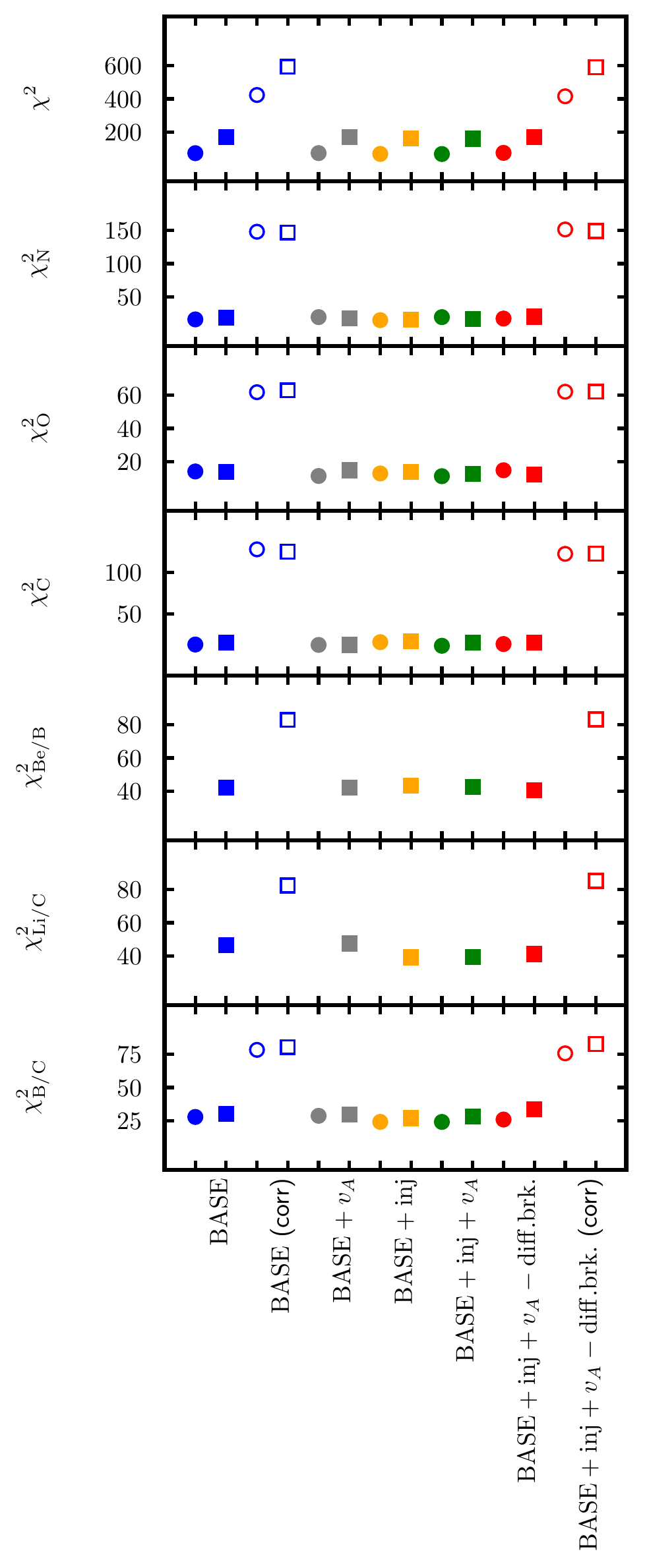}}
 \put( 0.24 , -0.0){\includegraphics[width=0.25\textwidth ]{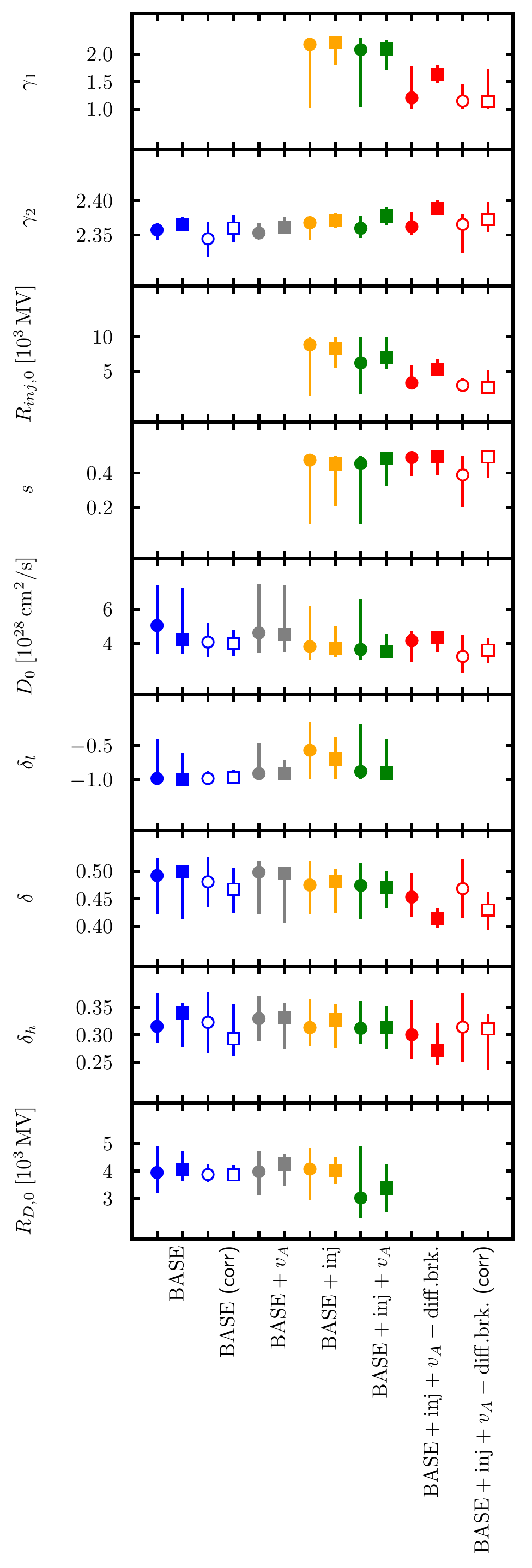}}
 \put( 0.505, -0.0){\includegraphics[width=0.25\textwidth ]{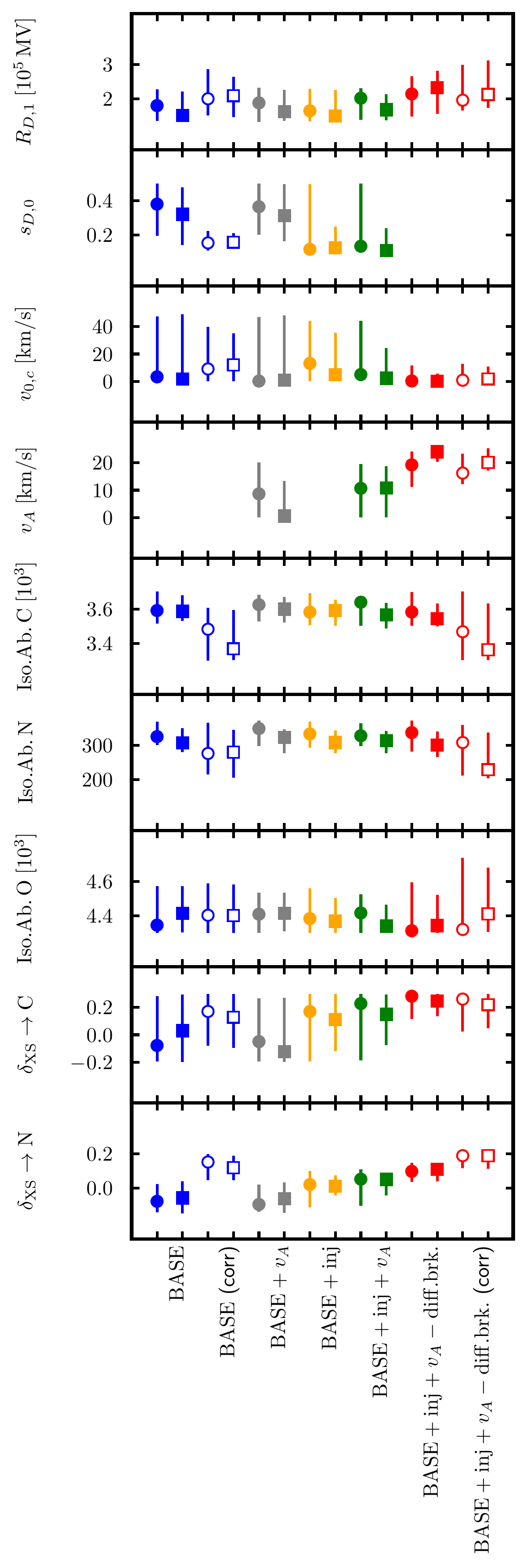}}
 \put( 0.77 , -0.0){\includegraphics[width=0.25\textwidth ]{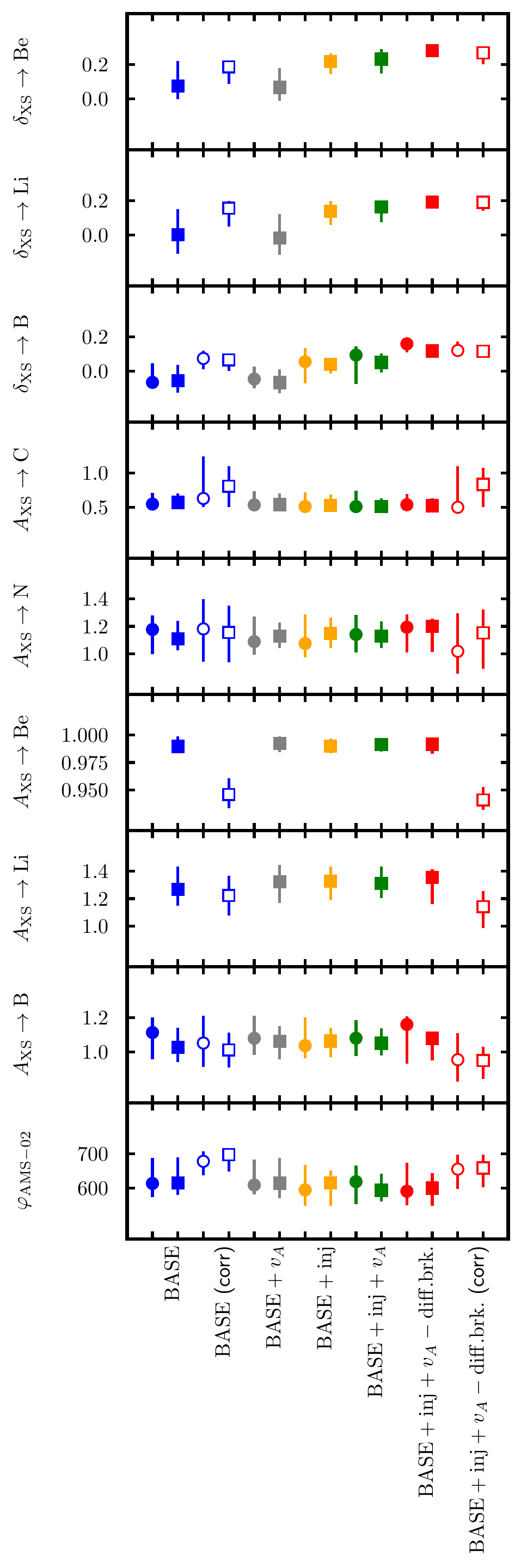}}
\end{picture}
\caption{ 
          For each CR propagation, cross-section nuisance, and solar modulation parameter we compare the best-fit value for the different CR propagation frameworks and 
          data sets. Circles refer to the BCNO data set, while squares mark the LiBeBCNO data set. Filled data points refer to our fiducial case with  uncorrelated AMS-02 systematic
          uncertainties. Open symbols in the cases  $\mathrm{BASE}$ (corr.) and  $\mathrm{BASE}+\mathrm{inj}+v_A-\mathrm{diff.brk}$ (corr.)  refer to the cases where the
          impact of correlated systematic uncertainties is considered. The error-bars display  2$\sigma$ uncertainties.
          \label{fig:Summary_param}
        }
\end{figure*}

\section{Results} \label{sec::results}

The results of the fits are summarized in Tab.~\ref{tab::fit_results}, which, 
for each fit, reports the total $\chi^2$ and the contribution to the $\chi^2$ for each single species, 
the number of degrees of freedom, the best-fit value for each parameter and the 1 $\sigma$ error. 
We can draw several  conclusions:

\begin{itemize}

\item
We can see that the fits are all very good with typical $\chi^2/\mathrm{dof}$ of about 0.3 for the BCNO 
fits and 0.45 for LiBeBCNO fits.
For better readability the same results are reported in graphical form in Fig.~\ref{fig:Summary_param}, 
while  Fig.~\ref{fig:BCNO_spectra} and Fig.~\ref{fig:LiBeBCNO_spectra}, instead, show the data for 
the spectra of each species together with the best-fit model predictions for each framework, 
with the bottom panels of each figure showing the residuals.
The quality of the fits can also be seen from the residuals which are all compatible 
with the size of the error bars themselves and thus at the level of 3-4\%.
Actually, the $\chi^2/\mathrm{dof}$ is a bit small, a result also seen in previous analyses~\cite{Korsmeier:2016kha,Cuoco:2019kuu}
and related to the fact that the systematic uncertainty reported by AMS-02 is probably slightly overestimated or correlated. 
While this typically indicates that the derived constraints should be conservative, 
one can also try to be more aggressive and model more explicitly the systematic uncertainty and its energy correlation, 
as we did for the four fits also reported in Tab.~\ref{tab::fit_results} and labeled ``(corr)''.  
One can see that indeed in these cases the $\chi^2/\mathrm{dof}$ increases to values a bit above one while the constraints on the parameters are slightly stronger.
A similar behavior is indeed also observed  in  Refs.~\cite{Cuoco:2019kuu,Heisig:2020nse}. 
For better clarity we do not display the residuals for the fits of correlated systematic uncertainties in Fig.~\ref{fig:BCNO_spectra} and~\ref{fig:LiBeBCNO_spectra}. 
As can be seen in the main panels, the best-fit lines of the fits with correlated uncertainties are often offset  with respect to the data points. However, this is expected and due to the fact 
that correlations in the data allow for an offset in normalization and/or small  tilts. Therefore, in these cases is more difficult or even misleading 
to judge the fit from the residuals and  one should  rather rely on the $\chi^2$ values themselves reported in the tables and Fig.~\ref{fig:Summary_param}. 

\item
We are able to obtain good fits with very similar best-fit $\chi^2$ for all the five framework explored. 
This means that the frameworks with fewer parameters should be preferred. For example, we see that the BASE framework in the BCNO case has a best-fit $\chi^2$ of 72.4. 
Adding one parameter more in the BASE+$v_A$ case leaves the $\chi^2$ basically unchanged. Adding three parameters more in the BASE+inj case improves the $\chi^2$ only by $\sim$ 4.5, while with four parameters more 
 in the BASE+inj+$v_A$ case  the $\chi^2$ improves only by $\sim$ 5.0. Using a likelihood ratio test, which in this case is equivalent to calculating the $\Delta \chi^2$, 
shows that all the improvements are at the level of about only 1 $\sigma$, and thus not significant. 
The BASE framework and the BASE+inj+$v_A$$-$diff.brk cannot be compared directly since they are not nested models. 
We can, however, compare the BASE+inj+$v_A$$-$diff.brk and BASE+inj+$v_A$ cases, which are nested. In this case,  
BASE+inj+$v_A$ has  three parameters more than BASE+inj+$v_A$$-$diff.brk and the $\chi^2$ is better by $\Delta \chi^2=7$. 
Based on the likelihood ratio and a $\chi^2$ distribution with 3 degrees of freedom, this is at the level of less than 2 $\sigma$, and thus again not significant. 
All in all, the two simplest models compatible with the data which emerge from this discussion are the BASE one and BASE+inj+$v_A$$-$diff.brk.  
Checking the $\chi^2$ values of the LiBeBCNO cases one can see that similar conclusions are reached. Also assuming correlated systematic uncertainties 
does not change these conclusions. 
Indeed, as can be seen from Fig.~\ref{fig:BCNO_propagation_triangle} which collects all the 2D contours related to the propagation parameters 
as well as the 1D profile $\chi^2$, the parameters of the BASE and BASE+inj+$v_A$$-$diff.brk  framework are generally well-constrained, 
while the more general framework BASE+inj+$v_A$, where all the parameters are considered, only appears to produce several degeneracies among the parameters and loose constraints, 
indicating, indeed, that the model has too many parameters with respect to the constraining power of the data.
These two models, as we already mentioned, are quite complementary and physically very different, since in the BASE case a break in diffusion 
is present at few GV which is not present in the BASE+inj+$v_A$$-$diff.brk  framework. This break is the crucial feature which in the BASE case 
allows us to fit the flattening and turning down observed at few GVs in the spectra of  secondary-over-primary ratios as can be seen in 
Figs.~\ref{fig:BCNO_spectra}-\ref{fig:LiBeBCNO_spectra}. At the same time, contrary to the BASE case, the BASE+inj+$v_A$$-$diff.brk  framework includes reacceleration, 
which is in this case the important feature allowing to fit the low-energy  secondary-over-primary ratios. 
It is quite remarkable that these two very different models can both provide good fits and are thus both valid models. 
Still, there is actually an important catch to this result regarding cross-section uncertainties, as discussed further below.

\begin{figure*}[!t]
\centering
\setlength{\unitlength}{1.0\textwidth}
\begin{picture}(1,0.85)
 \put(0.00, 0.35 ){\includegraphics[width=0.55\textwidth ]{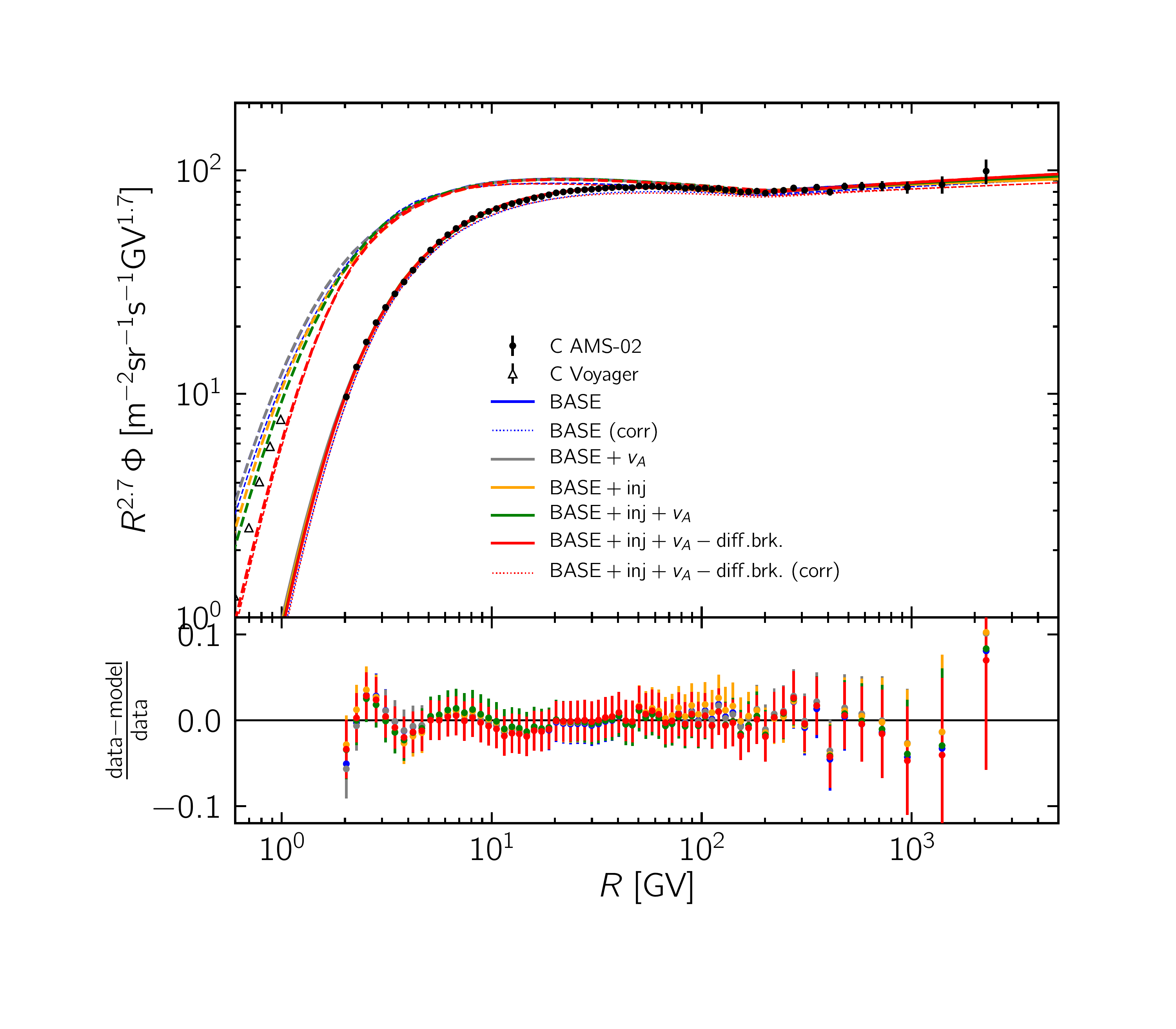}}
 \put(0.50, 0.35 ){\includegraphics[width=0.55\textwidth ]{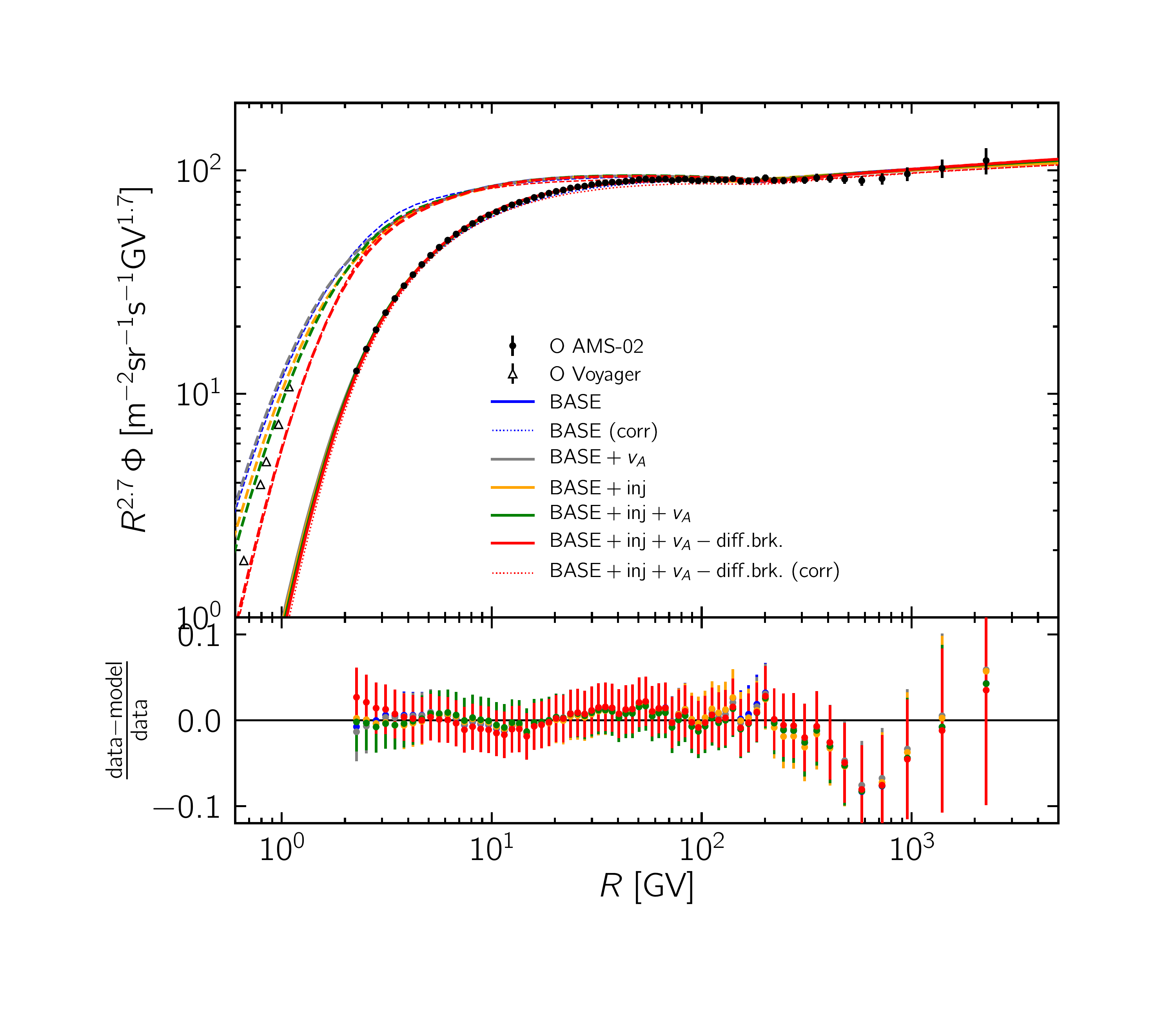}}
 \put(0.00, -0.05){\includegraphics[width=0.55\textwidth ]{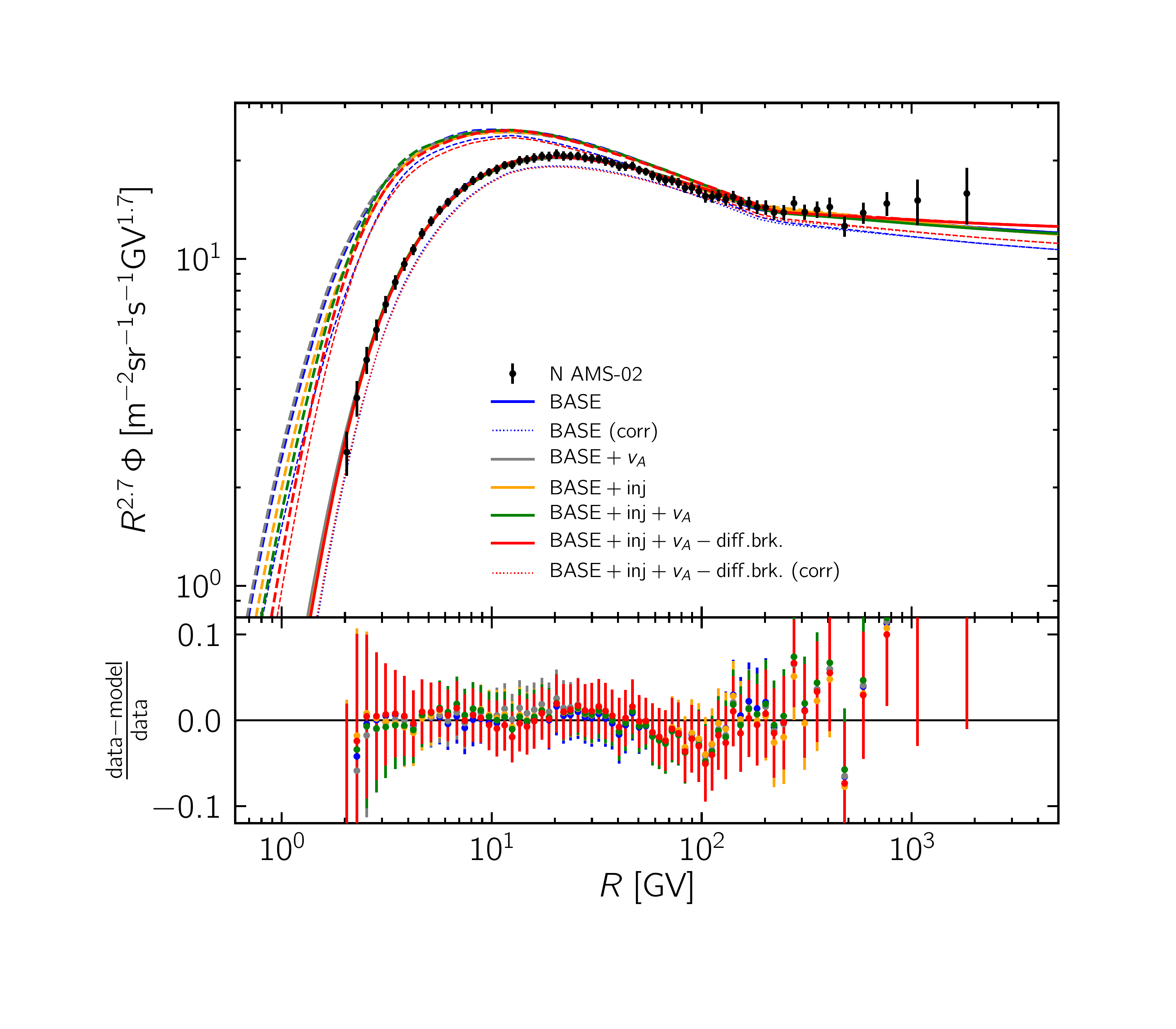}}
 \put(0.50, -0.05){\includegraphics[width=0.55\textwidth ]{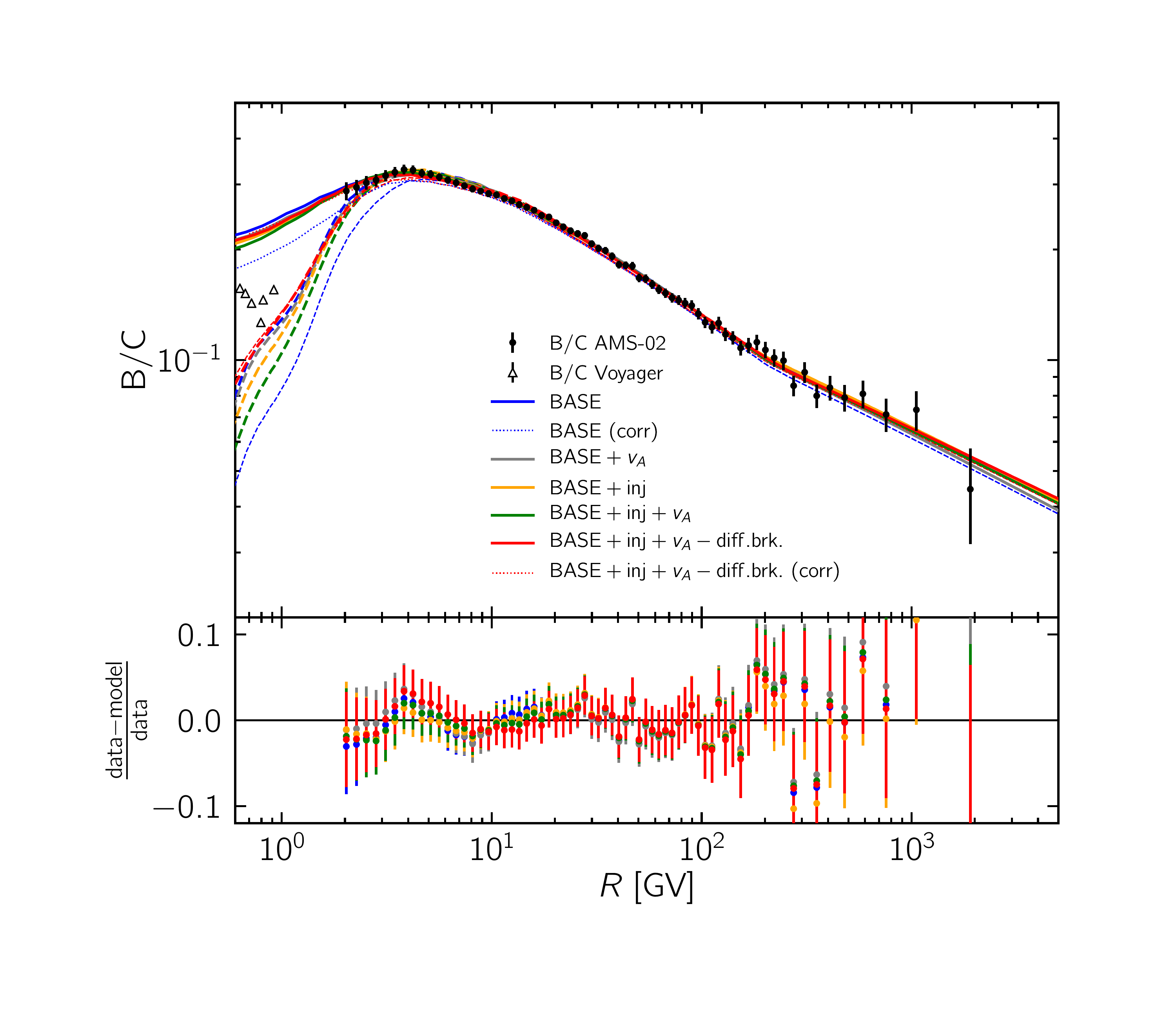}}
\end{picture}
\caption{ 
          Comparison of  AMS-02 measurements with the best-fit spectra for the BCNO data set. 
          The solid lines correspond to best-fit model for each of the five CR propagation frameworks. 
          The dotted lines are the best-fit models for the BASE and the BASE+inj+$v_A$$-$diff.brk scenario
          with correlated systematic uncertainties. 
          Dashed lines represent the interstellar CR spectra, \ie,\ the spectra before solar modulation is applied. 
          Residuals are only shown for the cases with  uncorrelated AMS-02 uncertainties. 
          For comparison we display Voyager data at low energies.
        }
\label{fig:BCNO_spectra}
\end{figure*}
\begin{figure*}[!t]
\centering
\setlength{\unitlength}{1\textwidth}
\begin{picture}(1,1.25)
 \put(0.00, 0.75 ){\includegraphics[width=0.55\textwidth ]{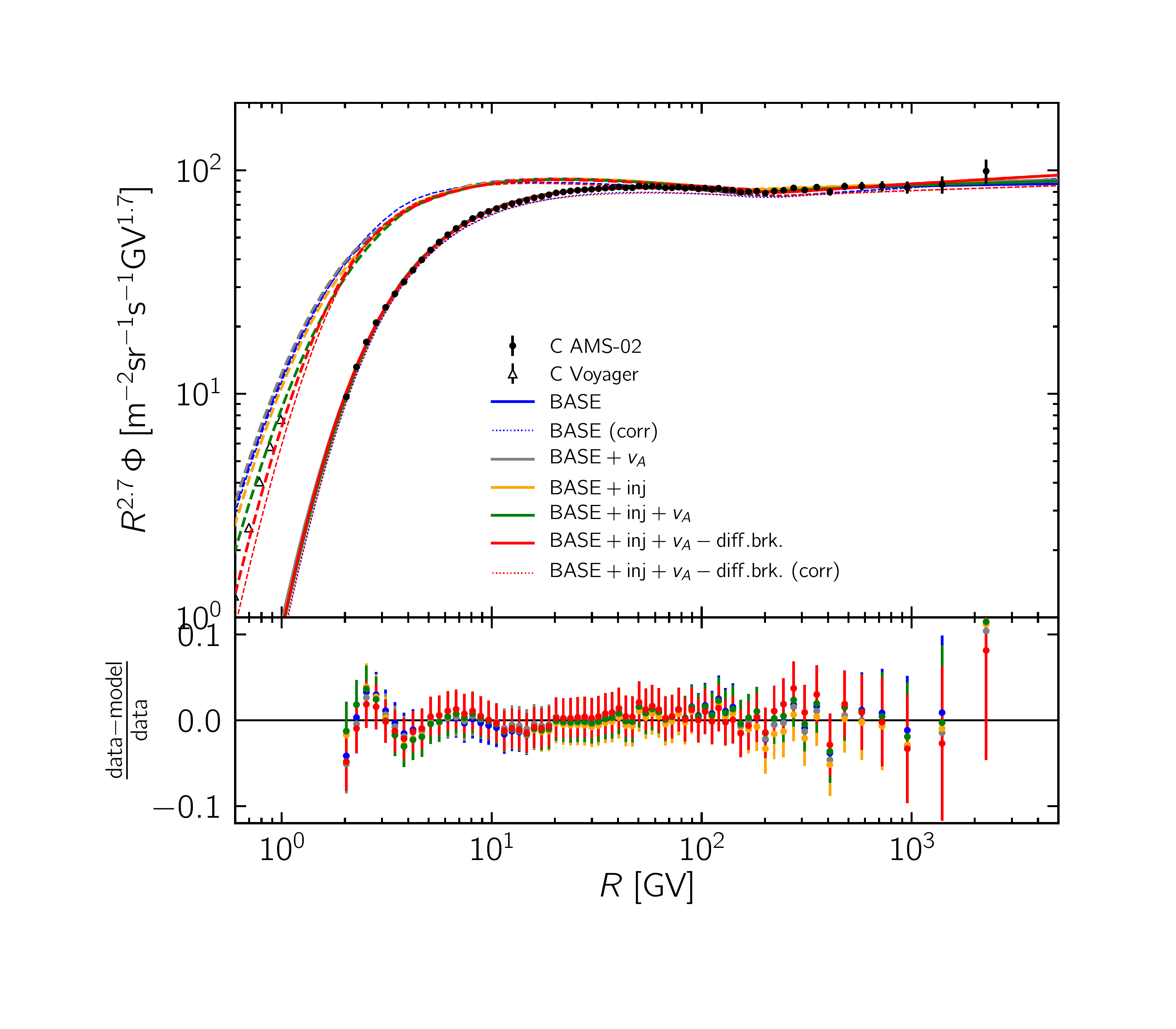}}
 \put(0.50, 0.75 ){\includegraphics[width=0.55\textwidth ]{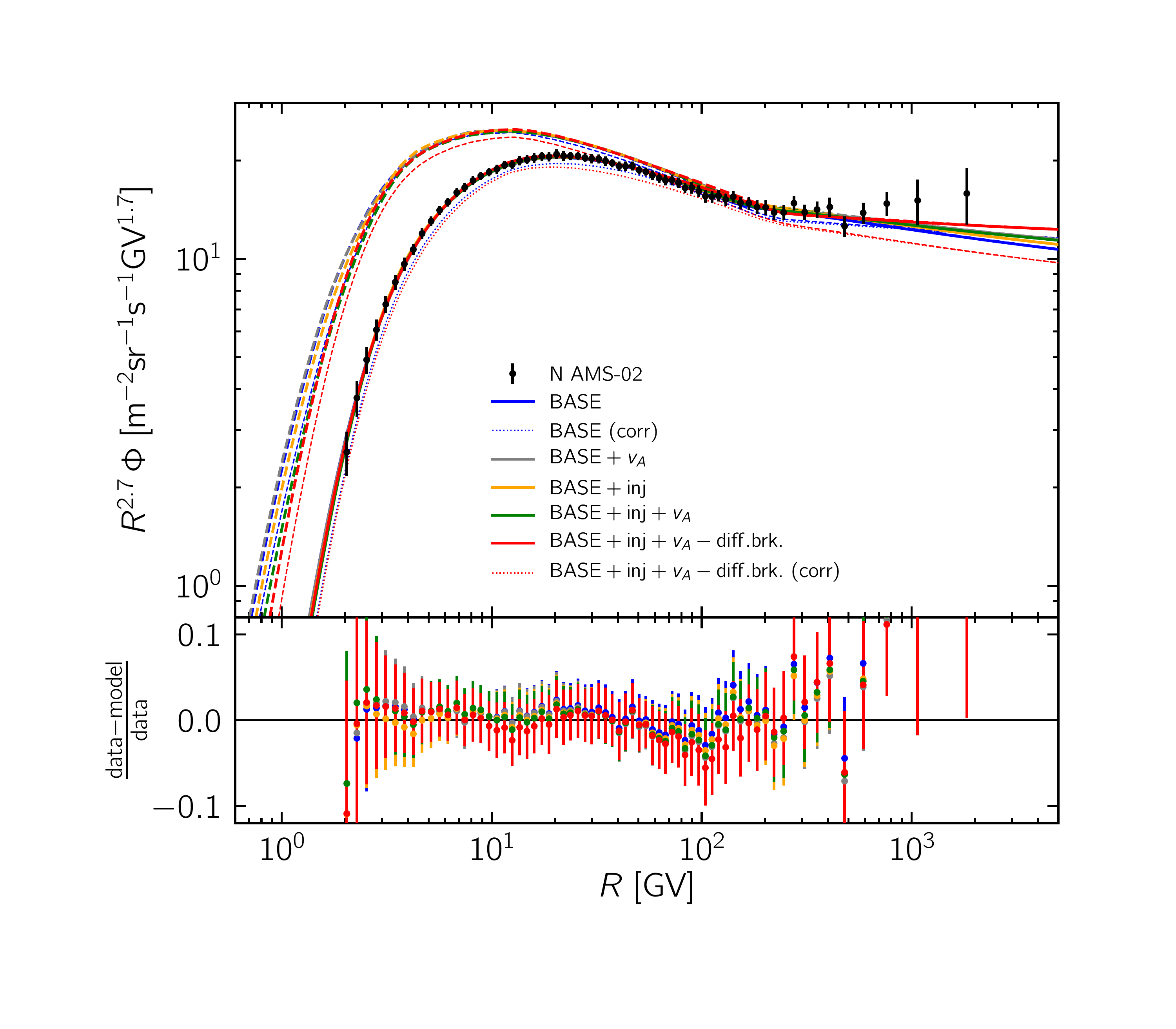}}
 \put(0.00, 0.35 ){\includegraphics[width=0.55\textwidth ]{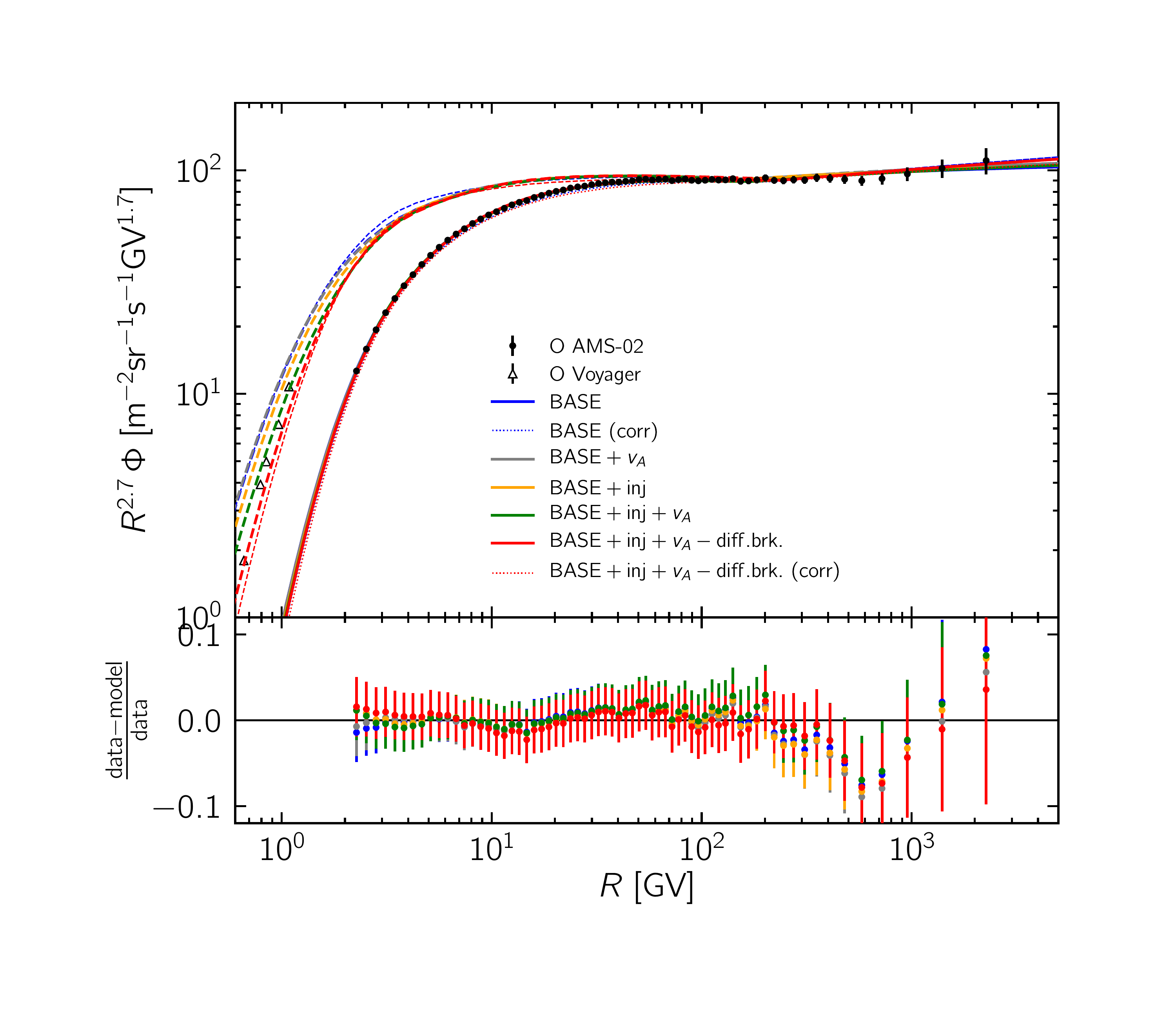}}
 \put(0.50, 0.35 ){\includegraphics[width=0.55\textwidth ]{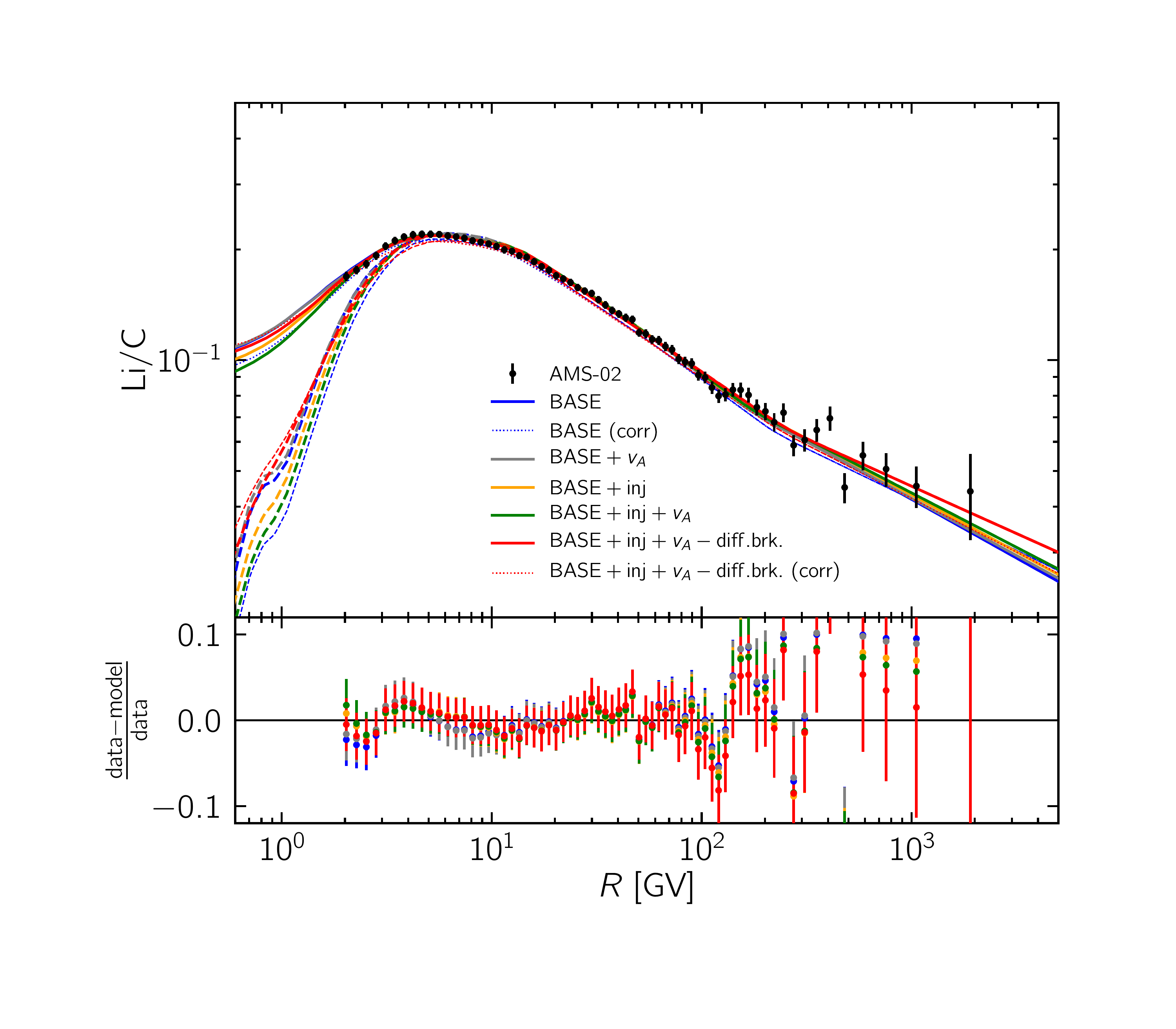}}
 \put(0.00, -0.05){\includegraphics[width=0.55\textwidth ]{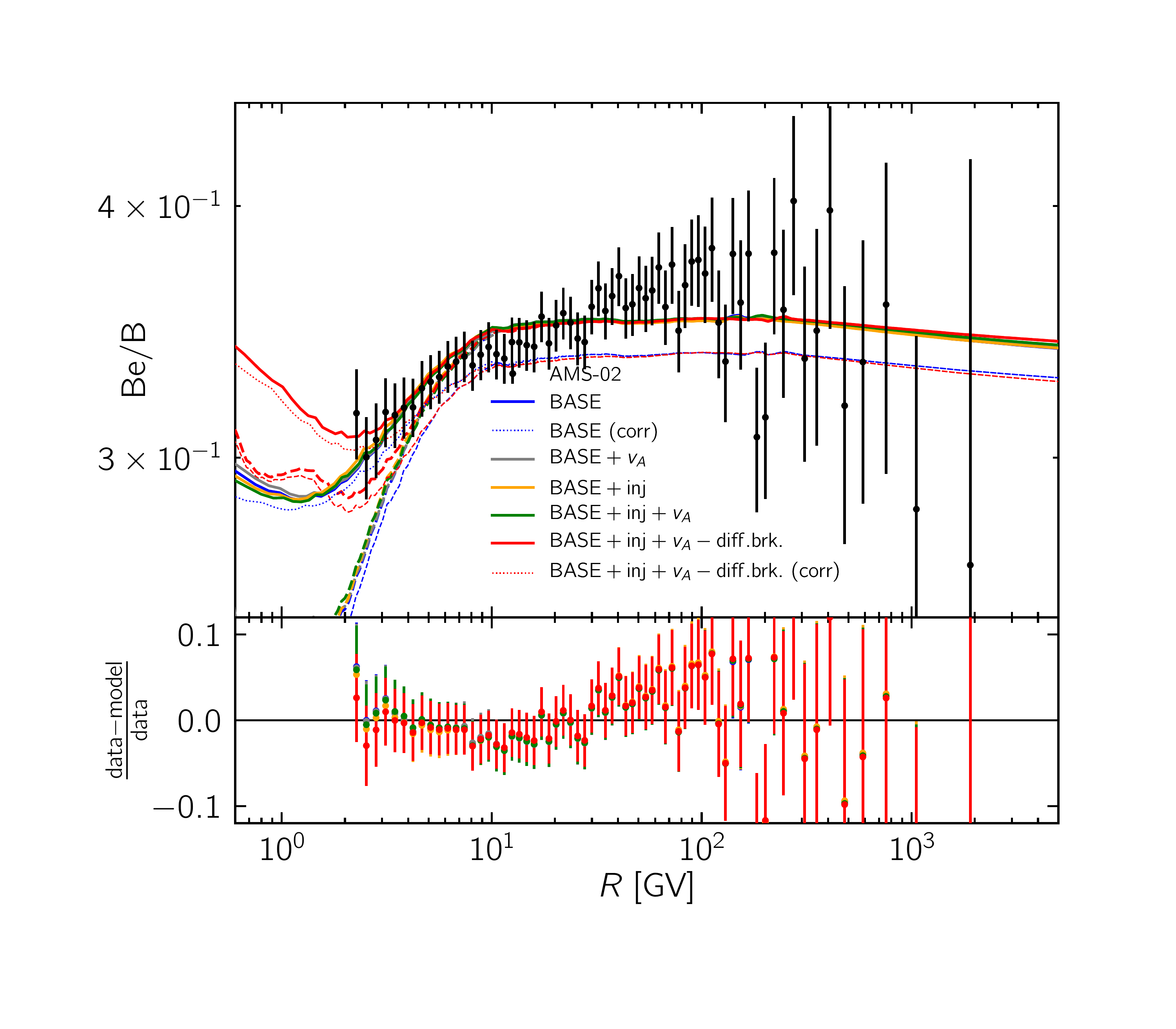}}
 \put(0.50, -0.05){\includegraphics[width=0.55\textwidth ]{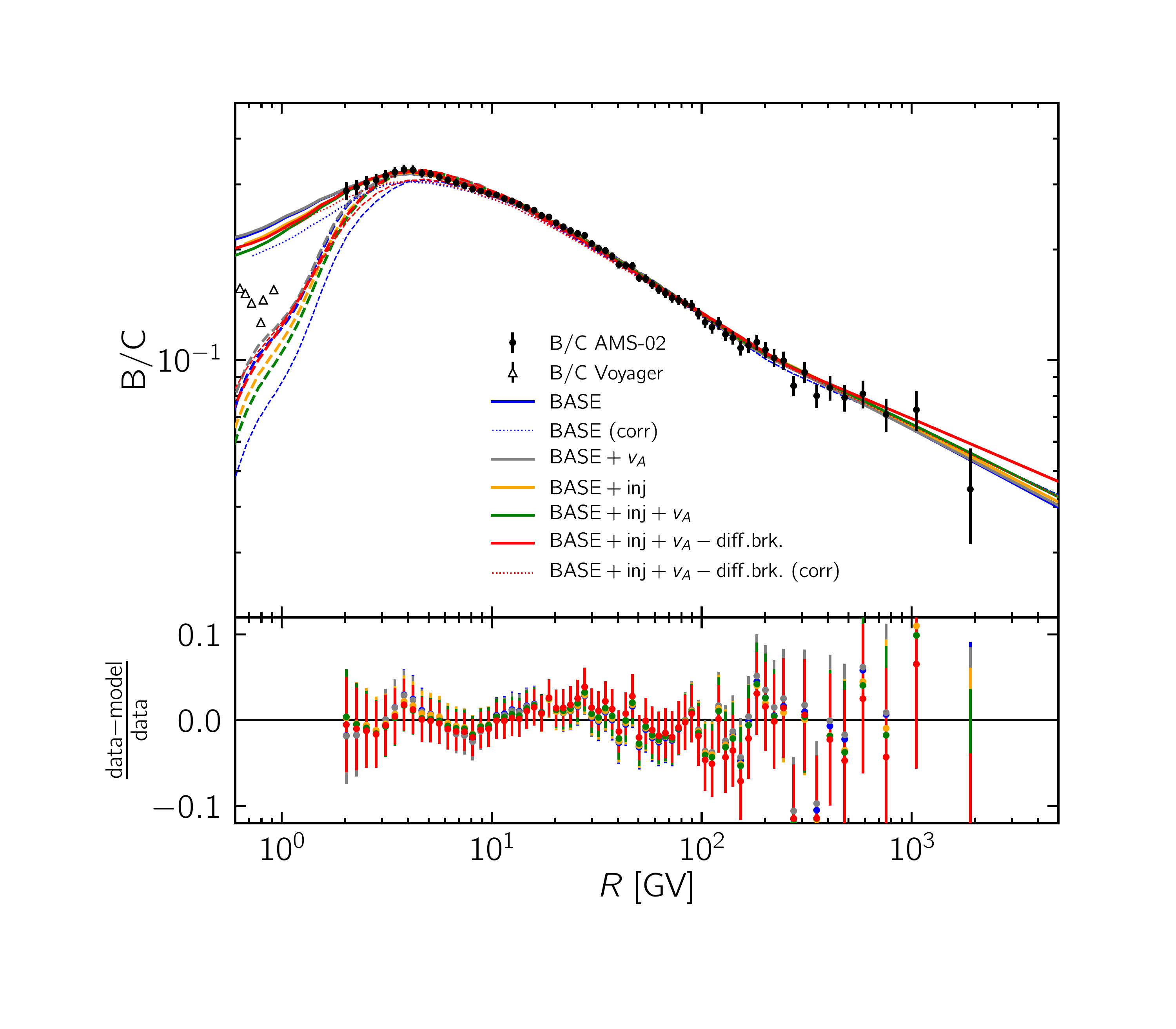}}
\end{picture}
\caption{ 
          Same as Fig.~\ref{fig:BCNO_spectra}, but for the fits with the LiBeBCNO data set. 
        }
 \label{fig:LiBeBCNO_spectra}
\end{figure*}
\begin{figure*}[!b]
\centering
\setlength{\unitlength}{1\textwidth}
\begin{picture}(1,0.51)
 \put(0.00, -0.0){\includegraphics[width=0.5\textwidth ]{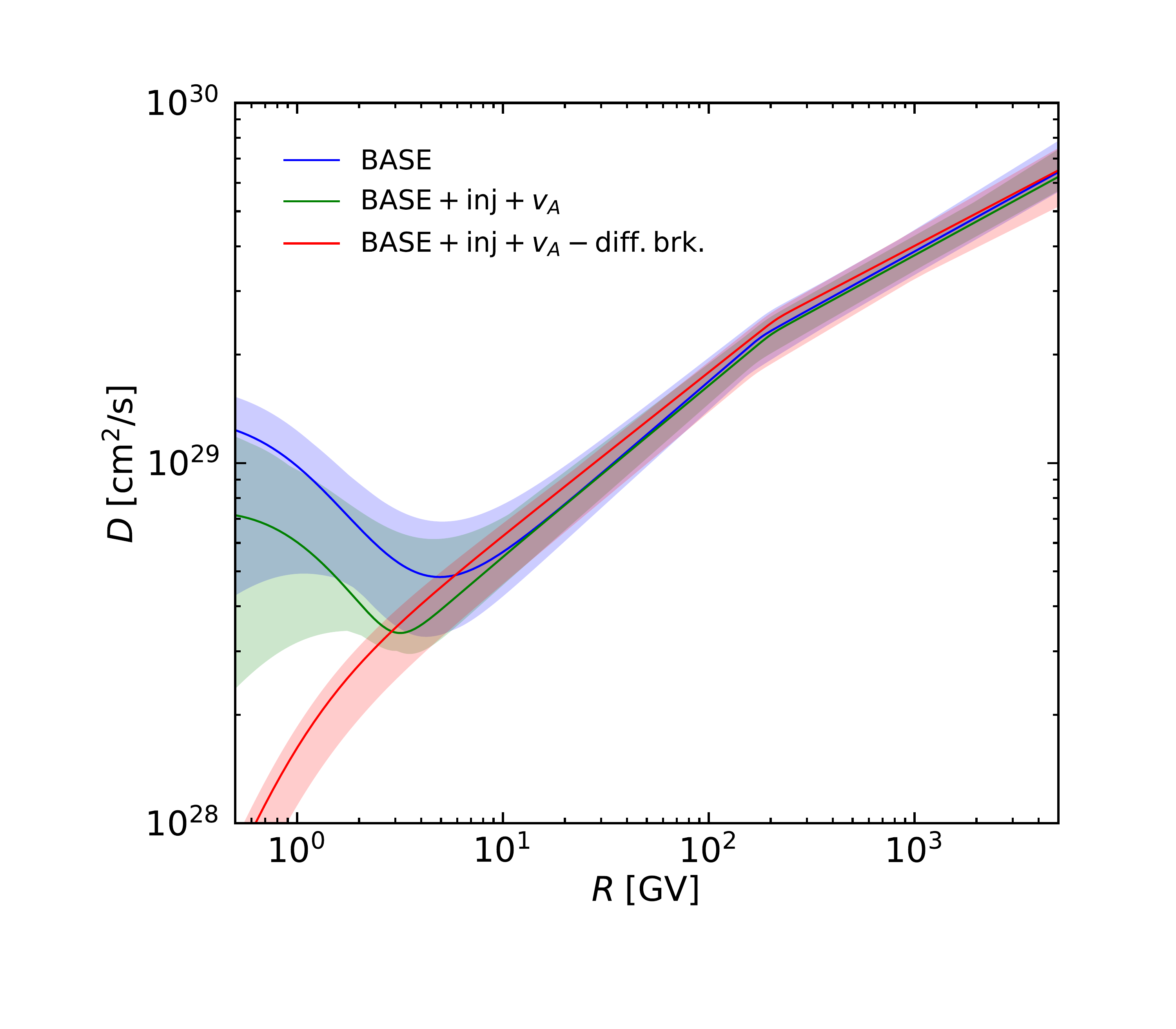}}
 \put(0.50 ,-0.0){\includegraphics[width=0.5\textwidth ]{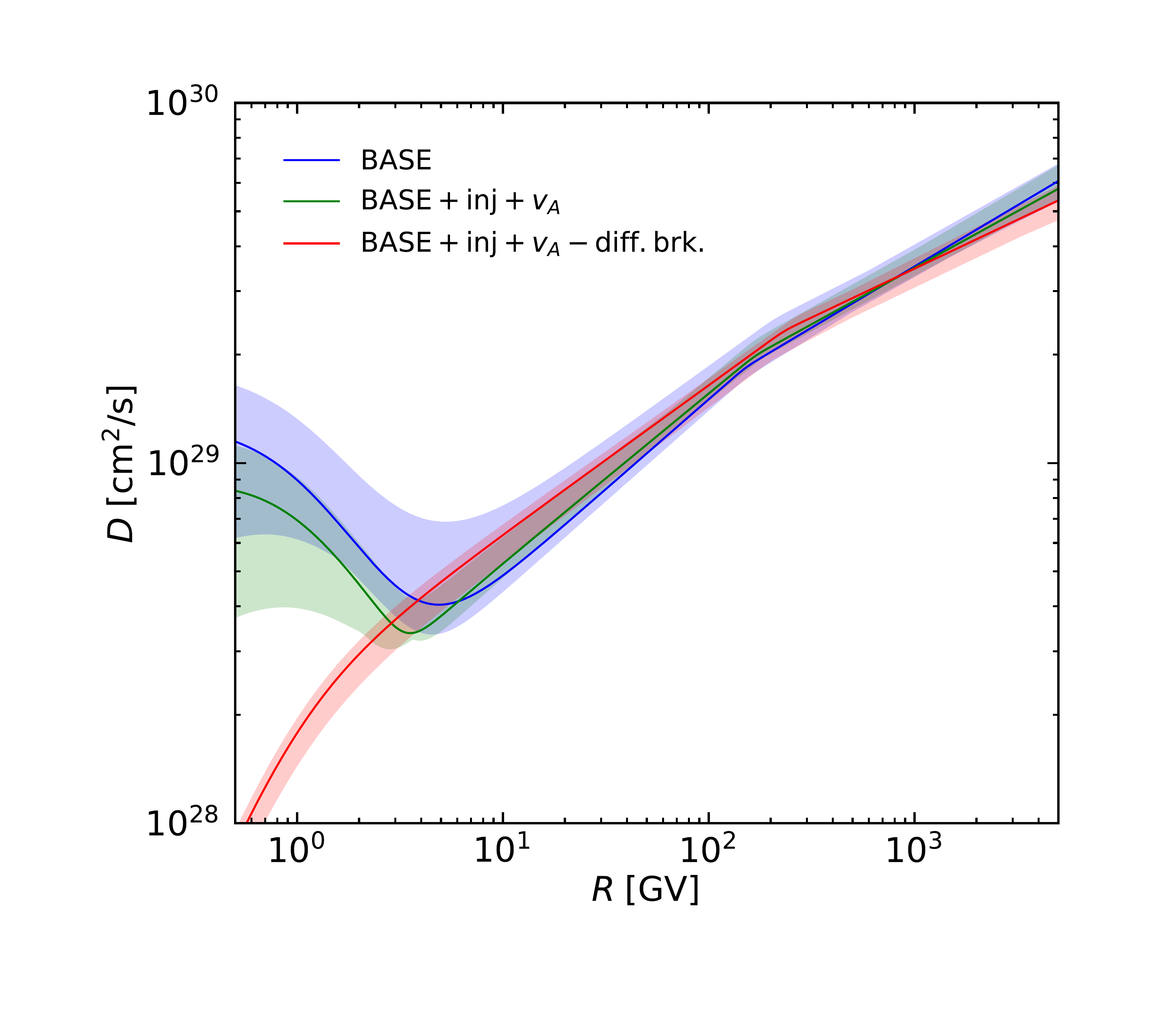}}
\end{picture}
\caption{ 
          Diffusion coefficient as function of rigidity for different fit configurations. The shaded band marks the $2 \sigma$ uncertainty. 
          In the fits the  half-height of the propagation region is fixed to a value of $z_\mathrm{h}=4\:\mathrm{kpc}$. 
          Left: fits with BCNO data. Right: fits with LiBeBCNO data.
        }
\label{fig:Dxx}
\end{figure*}
\begin{figure*}[t!]
\centering
\includegraphics[width=0.99\textwidth, trim={3cm 3cm 3cm 3cm},clip ]{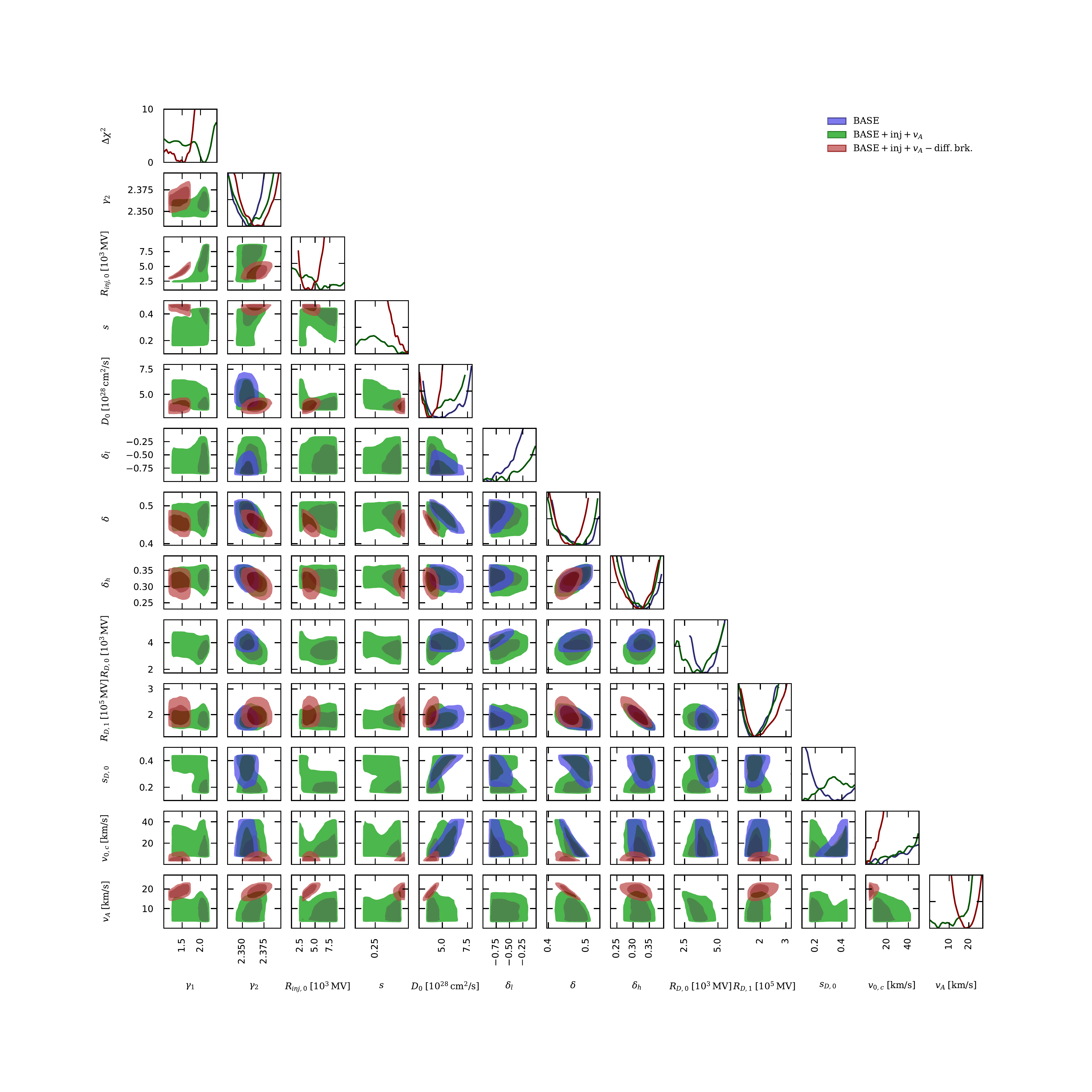}
\caption{ 
          Triangle plot with the fit results for the CR propagation frameworks  
          BASE (blue), BASE+inj+$v_A$ (green), and  BASE+inj+$v_A$$-$diff.brk (red), and using the BCNO data set. 
          The triangle plot shows only the subset of fit parameters related to CR propagation. 
          The 2D plots show the 1$\sigma$ and 2$\sigma$ contours for each combination of two parameters, which 
          are derived from the 2D $\chi^2$-profiles. The diagonal shows the 1D $\chi^2$-profiles (the $y$-axis is always the 
          $\Delta \chi^2$ and ranges from 0 to 10).
          \label{fig:BCNO_propagation_triangle}
        }
\end{figure*}
\begin{figure*}[t!]
\centering
\includegraphics[width=0.99\textwidth, trim={1.7cm 1.7cm 1.7cm 1.7cm},clip ]{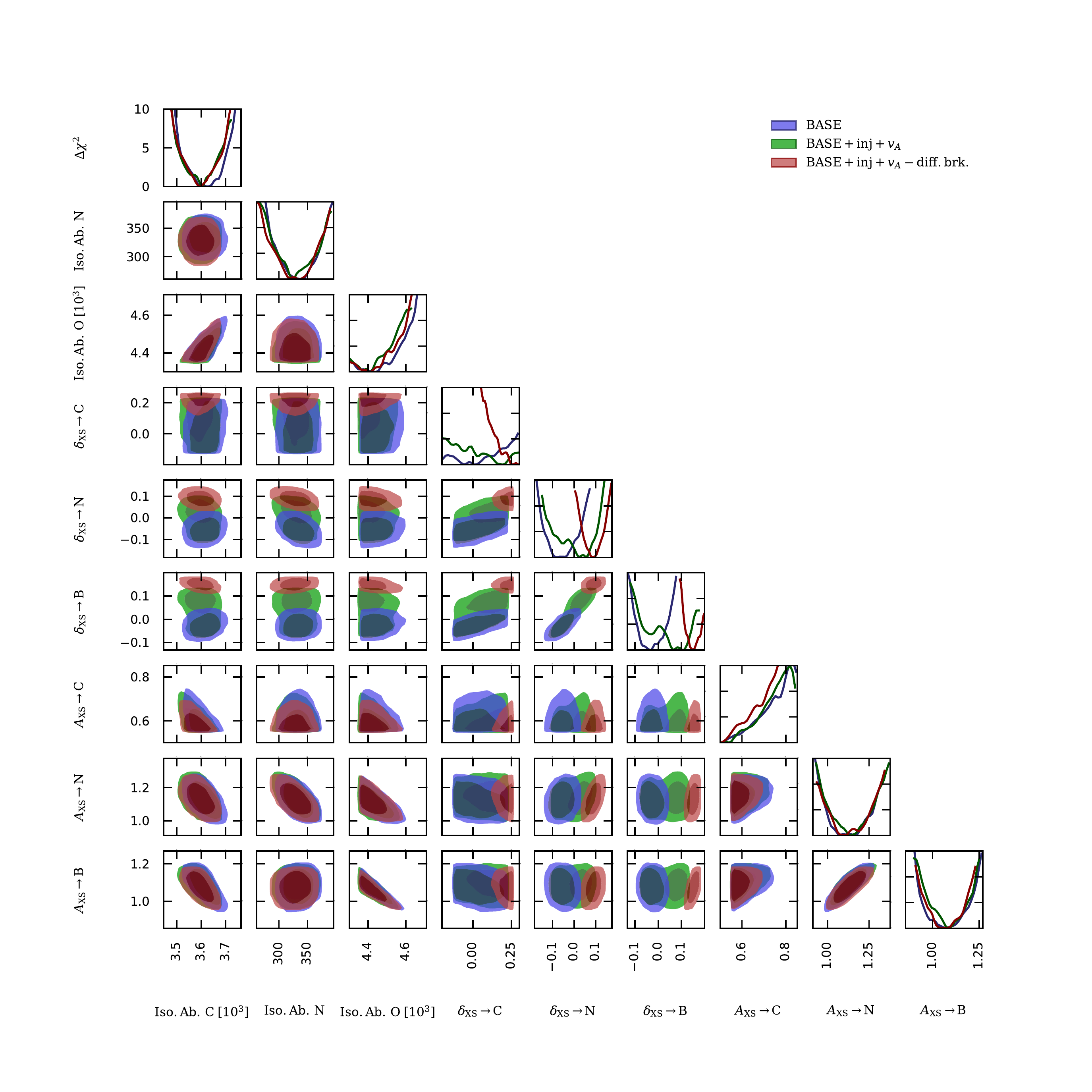}
\caption{ 
          Same as Fig.~\ref{fig:BCNO_propagation_triangle}, but for the subset of cross section nuisance parameters.
          \label{fig:BCNO_XS_triangle}
        }
\end{figure*}

\end{itemize}
\subsection{Results on  CR propagation parameters}
\begin{itemize}

\item
From Fig.~\ref{fig:Summary_param} we can see that all the propagation parameters are compatible in all the 14 fits, 
which is quite reassuring regarding the robustness of the result. There is one notable exception, 
regarding the $D_0-\delta$ plane which we will discuss in more detail below. Apart from this, 
it means that the three secondary nuclei Li, Be and B point to compatible propagation parameters 
and can be combined in a single global fit, a result which is certainly not trivial.

\item 
As mentioned above, the BASE and BASE+inj+$v_A$$-$diff.brk models point to compatible parameter values for all parameters they have in common.
At first glance, though,  there seems to be  a slight offset in the $D_0-\delta$ contours, as shown in the triangle plot in Fig.~\ref{fig:BCNO_propagation_triangle}.
However, this is mostly related to the normalization rigidity scale of the diffusion coefficient chosen at 4~GV, which is very close to the rigidity of the break in the BASE framework.
In Fig.~\ref{fig:Dxx} we show the diffusion coefficient as function of rigidity for the BASE, BASE+inj+$v_A$, and BASE+inj+$v_A$$-$diff.brk models.
Above 10\;GV the diffusion coefficients of all models agree well within the uncertainty. Furthermore, the diffusion coefficients derived from the BCNO and LiBeBCNO 
data sets match well, underlining again the compatibly of the three secondary CR species considered in this work.
We note that Fig.~\ref{fig:Dxx} is produced for a fixed $z_\mathrm{h}$ of 4\;kpc. If instead $z_\mathrm{h}$ were allowed to vary in the fits, the uncertainty band around the best-fit
diffusion coefficient would be enlarged significantly due to the well-known degeneracy between $D_0$ and $z_\mathrm{h}$. 

\item
We can see that for all of the fits convection is compatible with zero. This is actually due to a strong degeneracy between $v_{0,c}$, $D_0$, $\delta$ and, partly, $s_{D,0}$, 
at least for the models with a break in diffusion, as can be seen in Fig.~\ref{fig:BCNO_propagation_triangle}. 
For this reason, values of $v_{0,c}$ up to 50 km/s are still possible at the price of larger ranges allowed for $D_0$ and $\delta$.
The framework BASE+inj+$v_A$$-$diff.brk, instead,  prefers no convection allowing values only up to $\sim 10$ km/s.
At the moment, it is thus more economical to consider a model with no convection, since it fits the data equally well. 
In this case  lower values of $D_0$ are selected together with higher values of $\delta$ (since the degeneracy between $v_{0,c}$ and $\delta$ has a negative slope).
We note, nonetheless, that we are exploring only a simple model of convection with constant winds, 
but more complex models with wind gradients as function of $z_\mathrm{h}$ are possible and might give different results. 
A systematic study of convection goes beyond the aims of the current analysis.

\item
Due to the degeneracy discussed above for the BASE case, $\delta$ is constrained in the fairly large range 0.4-0.5 when convection is included, 
while assuming zero convection would give a preference for $\delta \simeq 0.5$  with a quite small error of the order of 0.01.
For the BASE+inj+$v_A$$-$diff.brk framework this degeneracy is weaker and the error on $\delta$ is thus smaller, but smaller values
of $\delta$ are preferred, namely $\delta \simeq 0.45$ and $\delta \simeq 0.41$ for the BCNO and LiBeBCNO fits, respectively.
We thus see that without convection the framework with reacceleration tends to prefer smaller values of $\delta$. 
These values of $\delta$ are in line with the results of \cite{Korsmeier:2016kha,Cuoco:2019kuu} where the \mbox{BASE+inj+$v_A$$-$diff.brk} model is analyzed, 
but with different CR data, namely proton, helium and antiproton. Also in this case, it is remarkable and not trivial 
that light nuclei and heavier nuclei point to a compatible propagation scenario. 
Other recent analyses \cite{Genolini:2019ewc} of AMS-02 data, which consider a break in diffusion, also tend to find higher values of $\delta$ 
with respect to a scenario with reacceleration and no break in diffusion.
From the point of view of diffusion theory this means that the latest AMS-02 data point to Kraichnan ($\delta\sim 0.5$) model of turbulence 
while Kolmogorov turbulence ($\delta = 1/3$) seems  disfavored.

\item
The models with a break in diffusion find a position of the break at around 4 GV, 
compatible with the predictions of theoretical studies~\cite{Ptuskin:2005ax,Blasi:2012yr}. The slope $\delta_l$ below the break is not well constrained.
There is a preference for a negative value, but, conservatively only an upper limit of about $\delta_l \lesssim 0$ can be derived. 
A negative value of $\delta_l$ is present, for example,  in models of turbulence with damping of Alfven waves at low energies~\cite{Ptuskin:2005ax}. 
All of the frameworks include a second break in diffusion at higher energies  and this is robustly detected in all of the fits, 
with a break around 200 GV and amount of breaking of about $\Delta \delta \sim 0.15$. 
This  confirms the findings of the AMS-02 analysis \cite{Aguilar:2018njt} where this result is derived 
modeling directly the observed spectra rather than from a diffusion model. It is also in agreement with the results of \cite{Genolini:2019ewc}. 
In absolute value the slope of diffusion at high energy becomes $\delta_h \sim 0.33$, which seems to indicate a transition 
from Kraichnan turbulence to Kolmogorov turbulence at few hundreds GVs.

\item
Reacceleration is a common ingredient in CR propagation models. However, typically, the amount of reacceleration required  to fit the CR data
is fairly large and might be in disagreement with energy considerations, as discussed recently in \cite{Drury:2016ubm}.
It is thus important to understand whether CR spectra do indeed require reacceleration for a good fit.
In the BASE+inj+$v_A$$-$diff.brk framework reacceleration is quite tightly constrained with a value of about 20 km/s. 
The BASE framework does not include reacceleration but is nonetheless able to fit the data well.
The more general framework BASE+inj+$v_A$ where both reacceleration and break in diffusion are included also allows values of $v_A$ up to 20 km/s, 
but due to the presence of degeneracies, $v_A$ is also compatible with zero. To understand if a certain amount of reacceleration is 
indeed physically present would thus require us to demonstrate the model BASE+inj+$v_A$$-$diff.brk  is preferred over the ones without reacceleration. 
However, at the moment, both models provide a good fit to the data, thus a definitive conclusion  is not possible yet.

\item
The abundances of primaries C, N, O are all well-constrained in the fit. There is, nonetheless, an expected degeneracy between the C and O normalization.
No clear correlation between the abundances of the primaries at the source and CR propagation parameters is observed,
although it would be expected at some level.
This is probably due to complicated degeneracies in the high-dimensional parameter space, which blur single 2D correlations. However, 
we note that there is an anti-correlation with the B production cross-section $A_\mathrm{XS} \rightarrow \mathrm{B}$. 
The values of C and O abundances preferred by the fit are about 20-30\% larger than typical values used in \texttt{Galprop}~\cite{Strong:2001fu}.  While the N abundance is larger by a factor of $\sim 2$.  The result reflects the fact that the C and O  flux normalizations measured by AMS-02 are about 20-30\% larger than pre-AMS-02 determinations (e.g. \cite{Adriani:2014xoa}, \cite{Aguilar:2017hno} and references therein).
These abundances can be also compared with measurements from the Sun  or meteorites \cite{Lodder:2003zy,Anders:1989zg}.
In this respect the O/N$\sim 10$ ratio found here is very similar to the value observed in the Solar system  given in \cite{Lodder:2003zy,Anders:1989zg}.

\end{itemize}
\subsection{The role of cross-section uncertainties}
\begin{itemize}

\item
A crucial role is played by the cross-section uncertainties that we have included in the fits. The triangle plots containing  
the 2D contours and 1D $\chi^2$ profiles of the relevant parameters are shown in Fig.~\ref{fig:BCNO_XS_triangle} (and also for the LiBeBCNO in the appendix: Fig.~\ref{fig:LiBeBCNO_XS_triangle}).
One can see that in all of the frameworks containing a break in diffusion, and thus in particular in the BASE framework, 
all of the cross-section parameters are compatible with having no deviations from the benchmark cross-section model (with a couple of notable exceptions to be discussed better below), 
\ie, the normalizations $A$ are compatible with 1 and the slopes $\delta_\mathrm{XS}$ are compatible with 0.  
Thus, this means that the CR spectra can be fit while only assuming the benchmark cross-section model, 
which is in this case  the \texttt{Galprop} one, without any modification. 
The important point to notice is that this is 
not valid for the reacceleration model BASE+inj+$v_A$$-$diff.brk. In this case significant deviations are seen. 
For example, the slope for B production needs to be modified to a value 
$\delta_\mathrm{XS} \rightarrow \mathrm{B} \simeq 0.15$ and it is incompatible with zero. 
Results  are similar for the other slopes: 
$\delta_\mathrm{XS} \rightarrow \mathrm{N} \simeq 0.1$, $\delta_\mathrm{XS} \rightarrow \mathrm{Be} \simeq 0.25$, 
and $\delta_\mathrm{XS} \rightarrow \mathrm{Li} \simeq 0.2$. 
The normalization parameters $A$, on the other hand, are compatible among all of the models.
The conclusion is that it would not have been possible to achieve a good fit for the BASE+inj+$v_A$$-$diff.brk framework 
if cross-section uncertainties were not included in the analysis. 
However, we cannot conclude that this favors the BASE model, since 
these deviations are only at the level of 20\% and are  within the systematic uncertainties present in the cross sections. 
Thus, the conclusion is rather the fact that the cross-section uncertainties are at the moment significantly limiting 
the inference of ISM propagation properties from CRs. If the uncertainties were much smaller, at the level of few percent, 
\ie, at the same level as the AMS-02 uncertainties, we would be able to select a specific propagation scenario over the others. 
As also discussed in other works~\cite{Genolini:2018ekk,Evoli:2019iih,Weinrich:2020cmw,Weinrich:2020ftb,Boschini:2019gow,Luque:2021nxb,Donato:2017ywo,Korsmeier:2018gcy}, 
this further confirms that a community effort is needed in order 
to carry out specific campaigns of measurements to reduce nuclear cross-section uncertainties, in such a way to have then the 
possibility to exploit the full potential of CR data to study the properties of the ISM. 

\item
As mentioned above, some deviation of the cross-section parameters from the benchmark model is present in all of the frameworks explored. 
The most evident case is the normalization of the cross-section for the secondary production of carbon, $A_\mathrm{XS} \rightarrow \mathrm{C}$,   
which always saturates the lower bound of our prior, \ie, a value of 0.5. 
Although uncertainties in the cross sections are typically quite large, a downward deviation at the level of 50\% 
or more  of the cross sections involving the production of carbon, in particular the ones coming from fragmentation of oxygen, seems unlikely.  
This deviation thus seems to indicate some underlying problem in fitting carbon. A possible solution is to allow different injections for carbon and oxygen. 
We verified with a specific fit that in this case $A_\mathrm{XS} \rightarrow \mathrm{C}$ becomes indeed again compatible with 1, 
and the fit gives slightly different injections for C and O. A different injection for  C and O may be justified 
in the light of the fact that also H and He are already observed to give indications of different injections. We stress, nonetheless, that in both cases 
this has a negligible impact on the propagation parameters, which is the main focus of this analysis.

Another parameter which presents a systematic deviation from the benchmark in all the fits is the normalization of the Li production cross section, 
$A_\mathrm{XS} \rightarrow \mathrm{Li}$, with a preferred value of $\simeq 1.3\pm0.15$. 
In this case, however, the deviation is within the expected uncertainties and thus acceptable. 
Similarly, although less pronounced, a systematic deviation is present in $A_\mathrm{XS} \rightarrow \mathrm{N}$ 
with a value of $\simeq 1.1\pm0.05$, again perfectly acceptable.

\end{itemize}

\section{Constraints on the Halo Half-Height $\bf z_\mathrm{h}$} \label{sec::zh}

In the previous section we have assumed a fixed halo half-height of $z_\mathrm{h}=4\;\mathrm{kpc}$, 
which is justified because of the  well-known degeneracy between $z_\mathrm{h}$ and $D_0$. 
As discussed in the introduction, this degeneracy can be broken with the use of \textit{radioactive clocks} (most notably $^{10}_{\phantom{1}4}\mathrm{Be}$). 
Measurements of $^{10}_{\phantom{1}4}\mathrm{Be}$ from AMS-02 are not yet available, however, the total Be spectrum measured by AMS-02 has 
small error bars at the level of few percent. The contribution of the $^{10}_{\phantom{1}4}\mathrm{Be}$ isotope to the total Be spectrum is at a level between 
10\% and 20\%. So, given the small uncertainties of AMS-02 measurement, the total Be spectrum might be able to break the 
$D_0$--$z_\mathrm{h}$ degeneracy.
In this section we thus explore the possible constraints on $z_\mathrm{h}$. To this end we  perform a total of 8 additional fits
using the same setup as in the previous fits but with a further free parameter, namely $z_\mathrm{h}$, left free to vary in the range between 2\;kpc and 10\;kpc.
For these fits we only consider the BASE and the BASE+inj+$v_A$$-$diff.brk frameworks with the two data sets BCNO and LiBeBCNO.
Furthermore, we consider both the case with uncorrelated uncertainties and the case with correlated ones. In total this gives eight different fits to explore.

\begin{figure*}[!t]
\centering
\setlength{\unitlength}{1\textwidth}
\begin{picture}(1,0.51)
 \put(0.00, -0.0){\includegraphics[width=0.5\textwidth ]{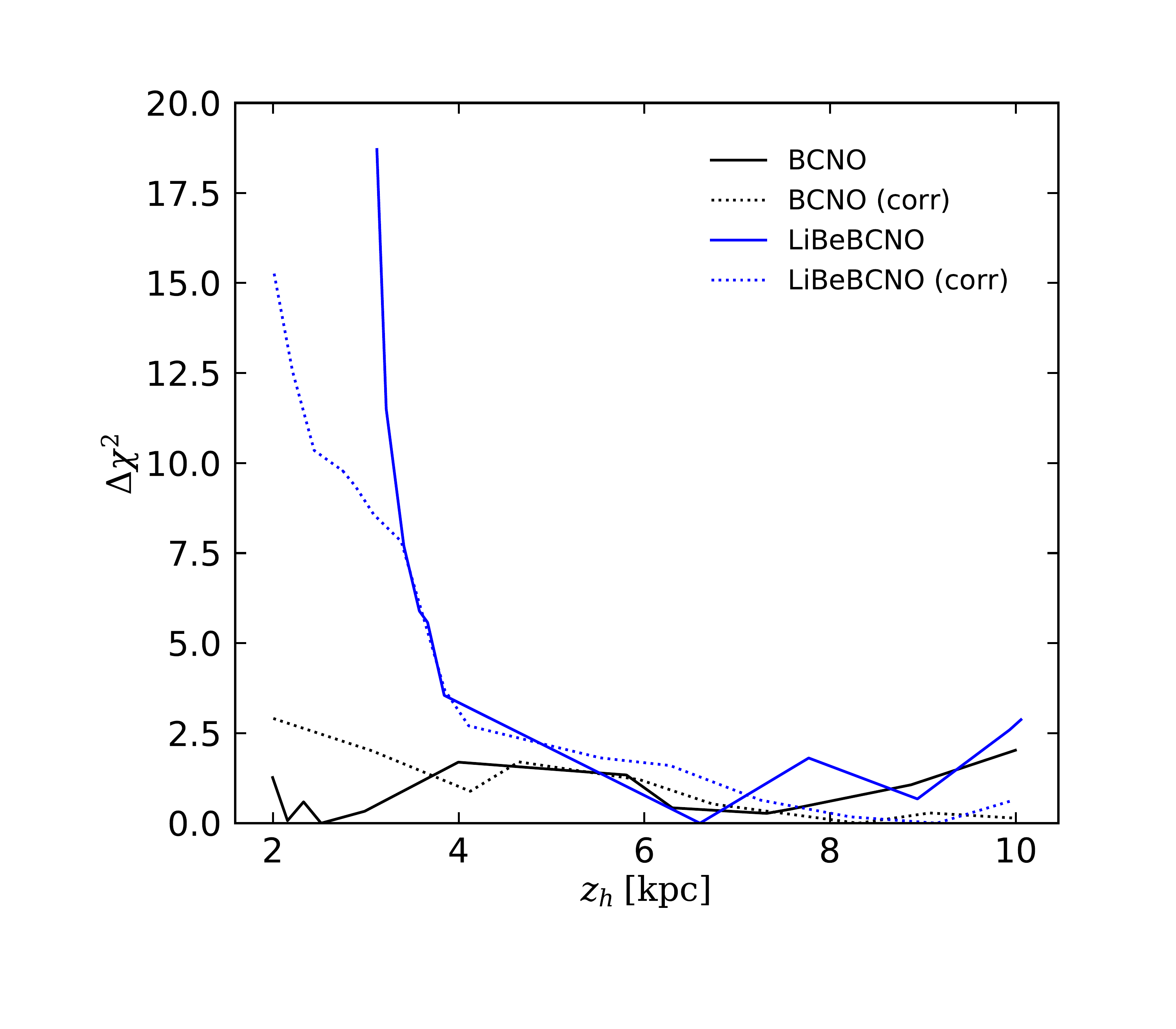}}
 \put(0.50 ,-0.0){\includegraphics[width=0.5\textwidth ]{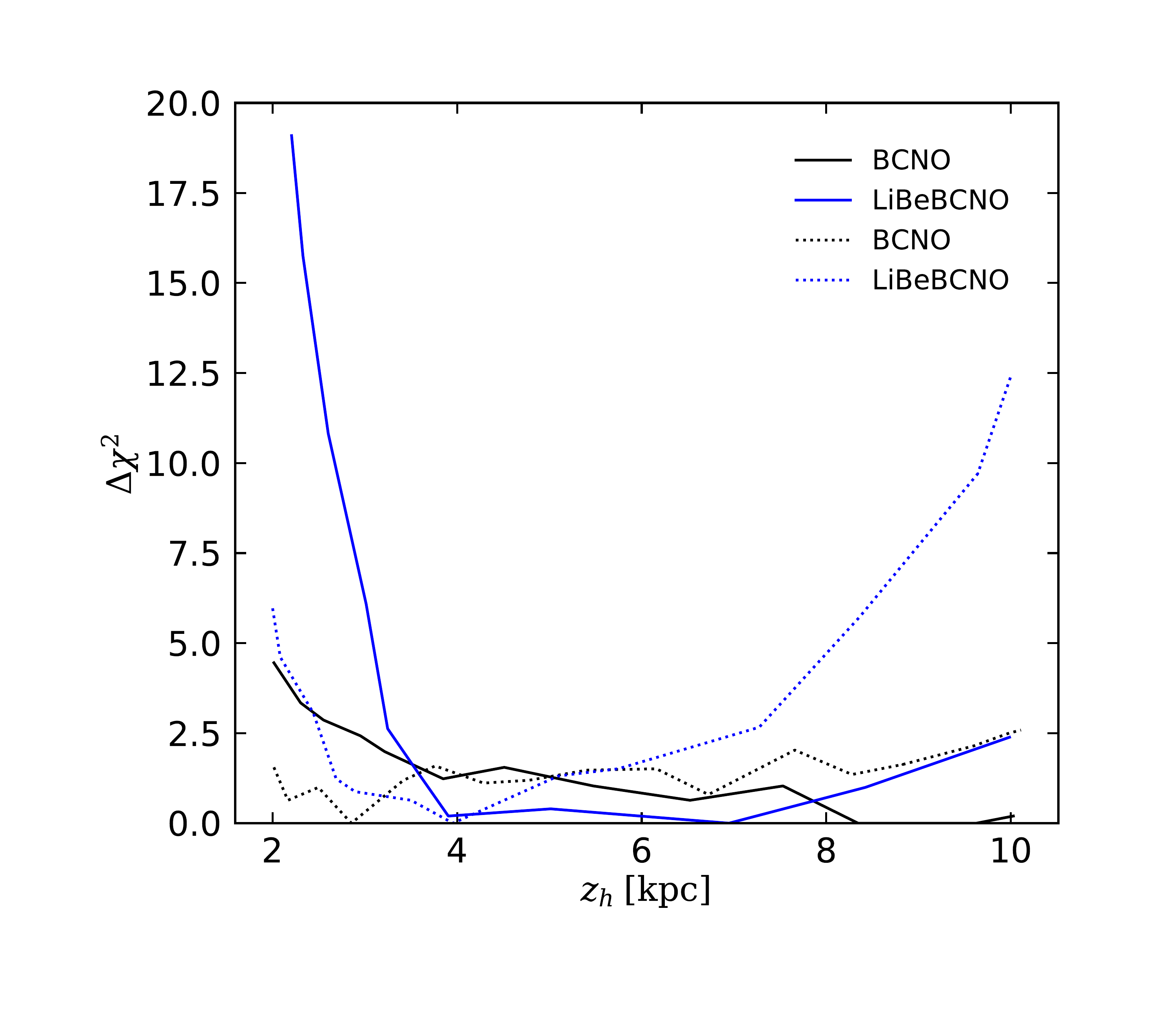}}
\end{picture}
\caption{ 
        	Constraints on the halo half-height $z_\mathrm{h}$. 
			The left panel refers to the BASE  framework, while the right one refers to the BASE+inj+$v_A$$-$diff.brk framework.
        }
 \label{fig:zprofile}
\end{figure*}

The $\chi^2$-profiles as function of $z_\mathrm{h}$ for all fits are shown in Fig.~\ref{fig:zprofile}. As expected, 
the BCNO data set is not able to provide  constraints on the halo height, and the $\chi^2$-profiles have a flat behavior and do not rise 
above the  2$\sigma$ level both for the BASE and BASE+inj+$v_A$$-$diff.brk scenario. 
The picture changes with the LiBeBCNO dataset, \ie, when the Be spectrum is included also in the fits. In the BASE
scenario (left panel of Fig.~\ref{fig:zprofile}) a robust lower limit on $z_\mathrm{h}\gtrsim 4\;\mathrm{kpc}$ at 2$\sigma$ (\ie, $\Delta \chi^2=4$)
can be derived, which is valid both when correlations in the data are considered and when they are not included.
No upper bounds on the halo height can be placed.
The situation is slightly different for the BASE+inj+$v_A$$-$diff.brk scenario. With uncorrelated uncertainties a lower bound on $z_\mathrm{h}\gtrsim 3\;\mathrm{kpc}$ at 2$\sigma$ is still present,
which is a bit less stringent than the BASE case. With correlated uncertainties, however, the lower bound appears to further weaken going down to 2 kpc. 
A weakening of the constraints when correlated uncertainties are considered is in principle possible since correlations introduce some freedom in the normalization 
and long rigidity-scale tilt in the data which can be somehow degenerate with the effect of varying some parameters, such as $z_\mathrm{h}$. Indeed, this effect seems to be present 
also in the BASE case, although in that case it seems to be less prominent.
Finally, an upper bound of  $z_\mathrm{h}\lesssim 8\;\mathrm{kpc}$ at 2$\sigma$ is given in the correlation case, which  also does not appear to be very robust since it is
not present in the uncorrelated case.
Overall, we conclude that the constraints on $z_\mathrm{h}$ appear to be both propagation-scenario dependent and dependent on the assumptions 
regarding the correlation properties of data systematics, and thus they should be taken with some care.
Nonetheless, from the above analysis we can infer a strong indication, although not a conclusive one, for a lower limit of about 
3 kpc on $z_\mathrm{h}$. The situation might likely change and become more constraining once AMS-02 data on the $^{10}_{\phantom{1}4}\mathrm{Be}$ isotope becomes available.

\section{Discussion and comparison with other works} \label{sec::comparison}

In this section we discuss and compare our results with previous works.
The analysis in  Refs.~\cite{Weinrich:2020cmw,Weinrich:2020ftb}, similarly to this work, takes into account  nuclear cross-section uncertainties.  
A main difference is the fact that we include primaries in the fit to constrain the diffusion framework 
while  Refs.~\cite{Weinrich:2020cmw,Weinrich:2020ftb} consider only secondaries for this purpose. 
Another difference is the use  of a simplified semi-analytical framework for propagation as opposed 
to our fully numerical treatment through the  \texttt{Galprop} code.
Nonetheless, overall our results are in reasonable agreement.  We both find a value of $\delta \sim 0.5$ in models without reacceleration 
and with a break in diffusion, and a tendency to get a smaller value of $\delta$  in the complementary case with reacceleration but without  break in diffusion.
A further common conclusion is that the inclusion of cross-section uncertainties is crucial to achieve a good fit 
and that sizable deviations from the benchmark models are observed. 
Ref.~\cite{Weinrich:2020ftb} constrains the $z_\mathrm{h}$ to $5\pm3\;\mathrm{kpc}$ at 1$\sigma$. The lower limit is in agreement with our findings.

In  Refs.~\cite{Evoli:2019wwu,Evoli:2019iih,Schroer:2021ojh} a simplified analytical approach similar to that of  Refs.~\cite{Weinrich:2020cmw,Weinrich:2020ftb} is used.  
They perform the fit only above 20 GV to exclude the rigidity region most strongly affected by Solar modulation.  
They do not include nuclear cross section uncertainties, although they employ an updated parametrization  from a refit of the relevant processes.  
They find a  constrain on $z_\mathrm{h}$ in the range 3-6 kpc  when using only the AMS-02 statistical error bars. 
When systematic uncertainties are included, the constraint is softened  to approximately $z \gtrsim 2$ kpc.  
They find a best fit for $\delta = 0.54$ (in Ref.~\cite{Evoli:2019iih} fitting to B/C, Be/C, B/O, and Be/O data) and
$\delta = 0.63$ (in Ref.~\cite{Evoli:2019wwu} fitting to B/C, C, N, and O data) 
which is in principle incompatible with our findings, although they 
do not report any error so a proper comparison is not possible. 

In Refs.~\cite{Boschini:2018baj,Boschini:2019gow} the authors use a propagation framework equivalent to the BASE+inj+$v_A$$-$diff.brk 
one together with a fully numerical treatment both for Galactic propagation using the \texttt{Galprop} code and 
for Solar System propagation using the Helmod code. Nuclear cross-section uncertainties are not included.
The main finding is the need to consider primary Li in order to achieve a reasonable fit for Li CR data.
This is possibly  related to the choice of not allowing freedom in the cross-section normalization. 
As we have shown, we are able to fit Li well but with some substantial deviation in the Li production 
cross section from the benchmark model. A similar deviation for the Li cross-section is indeed found
also in  Refs.~\cite{Weinrich:2020cmw,Weinrich:2020ftb} and  Refs.~\cite{Luque:2021nxb,Luque:2021joz}. Finally, they find a value of $\delta= 0.415\pm0.025$ which is compatible 
with the value $\delta= 0.41\pm0.01$ we find for our LiBeBCNO fit in the BASE+inj+$v_A$$-$diff.brk framework.

In Refs.~\cite{Luque:2021nxb,Luque:2021joz}, the authors study the CR spectra of Li, Be, and B also taking cross section uncertainties into account. 
They use the \texttt{Dragon2} code to model CR propagation in a framework similar to BASE+inj+$v_A$$-$diff.brk. To model cross section uncertainties they compare the two 
default cross section options of \texttt{Dragon2} and \texttt{Galprop} and, furthermore, they iteratively adjust the normalization of the secondary production 
cross sections within uncertainties to obtain a better fit to the AMS-02 data. They confirm that different models of cross sections have a significant impact on the final predicted CR spectra.
In agreement with our results, they also report a preferred normalization of the Li production cross section around $\sim$1.3 when using the \texttt{Galprop} model. 
As in Refs.~\cite{Weinrich:2020cmw,Weinrich:2020ftb} they use only secondary species to constrain propagation.
Regarding the slope of diffusion they find a value of $\sim 0.4$ in agreement with what we find for the BASE+inj+$v_A$$-$diff.brk case.

\section{Summary and Conclusions} \label{sec::conclusions}

In the present work, we have performed global fits on the recently published AMS-02 measurements of secondary nuclei
Li, Be, B, primaries C and O and mixed secondary/primary N, in order to study the propagation properties of CRs
in the ISM.  
We consider five different propagation frameworks differing with respect to the inclusion of a diffusion break at few GVs, the presence 
of reacceleration, and the presence of a break in the injection spectra of primaries.
The first one, which we dub BASE, has the injection spectrum of primaries modeled as a simple power law, no reacceleration
and a break in the diffusion coefficient at about few GVs in rigidity. 
The BASE+$v_A$ one, as the name suggests, enlarges the first one including reacceleration. Similarly,  BASE+inj
includes a break in the injection of primaries, while  BASE+inj+$v_A$ includes both of them and it is the most general model.
Finally, the model BASE+inj+$v_A$$-$diff.brk
has reacceleration,  a break in injection also at a few GVs, but no break in the diffusion coefficient.
In particular, among these, BASE and BASE+inj+$v_A$$-$diff.brk are the ``minimal'' two, in the sense that
they have the minimal number of parameters. Furthermore they are also the ones which differ the most from the physical point of view.
In the rest of the conclusions we will thus focus on discussing these two models.

It is well known, however, that the production cross section of secondary CRs are affected by sizable uncertainties 
and this can lead to a bias in the inferred propagation properties. To properly take into account this uncertainty
we consider  nuisance parameters in the cross sections in the form of overall normalizations and tilts in the low energy regime
below few GeVs. We thus extend our formalism in order to perform a global fit where both propagation parameters and
cross-section nuisances are included together so that to properly consider all these uncertainties and their  correlations.
We use Monte Carlo scanning techniques in order to handle the large ($> 20$) dimensionality of the parameter space. 
Finally, a further uncertainty we explore is the presence of correlation in the systematic error bars of AMS-02, a possibility which has 
been discussed recently.

The main result of our analysis is that all the five propagation scenarios we examine are able to provide good fits to the AMS-02 data,
including both the BASE and BASE+inj+$v_A$$-$diff.brk scenarios, which is remarkable, since, as stressed above, they describe quite
different physical scenarios.
Crucial to this result, however, is the fact that we have taken into account the cross section uncertainties.
While in the BASE scenario the preferred values of the cross-section nuisance parameters are in line with the default cross-section model we use, 
namely the \texttt{Galprop} model, in the BASE+inj+$v_A$$-$diff.brk scenario sizable deviations are required, especially in 
the tilt at low energy, at the level up to 20-30\%. However, these larger cross sections are currently
within the level of uncertainties of these cross sections, and thus acceptable.
Thus, the conclusion is that, at the moment, the nuclear cross section uncertainties are significantly limiting our capacity to
study the properties of the ISM and the propagation of CRs. In fact, since the AMS-02 data are in principle stringent enough to discriminate
among different scenarios, the cross section systematics are able to introduce enough uncertainty to make them degenerate. 
It is thus desirable to have  experimental efforts in order  to reduce these uncertainties with proper measurement campaigns in the laboratory.

Nonetheless, several conclusions can still be drawn which are robust to the inclusion of the cross-section uncertainties.
For example, the slope of the diffusion coefficient at intermediate energies is well constrained
in the range $\delta\simeq0.45-0.5$ in models without convection, or 
$\delta\simeq0.4-0.5$ if convection is included in the fit, pointing to a Kraichnan model of turbulence of the ISM.
We also find that convection is not required in any of the five considered CR propagation scenarios,
although we did not perform a systematic exploration of different convection models but we only considered a simple model with constant convective winds.
The BASE+inj+$v_A$$-$diff.brk requires reacceleration  with Alfven waves of about $v_A\simeq 20$ km/s. 
However,  since the BASE model does not require reacceleration and fits the data equally well, at the moment 
the data are inconclusive about the presence of reacceleration. This is an example of the physics
limitation introduced by the nuclear cross-section uncertainties.
Finally, since Be data are sensitive to the vertical size of the propagation halo $z_\mathrm{h}$, due to the contribution
of the radioactive isotope $^{10}_{\phantom{1}4}\mathrm{Be}$, in the fits including Be data we are able to derive a lower limit on $z_\mathrm{h}$
of about $z_\mathrm{h}\gtrsim 3\;\mathrm{kpc}$ at 2$\sigma$.
However, while all of the previous results are robust with respect to the inclusion of correlation in the AMS-02 data systematics
this last result gets weakened to  $z_\mathrm{h}\gtrsim 2\;\mathrm{kpc}$ when correlations are included,
especially in the case of BASE+inj+$v_A$$-$diff.brk. On the other hand, the BASE scenario provides a more robust and stringent lower limit
of $z_\mathrm{h}\gtrsim 4\;\mathrm{kpc}$. Given the dependence of these constraints on the propagation scenario
and the correlation properties of the data, the Be spectrum does not seem optimal to study $z_\mathrm{h}$. 
Thus, precise measurements of the $^{10}_{\phantom{1}4}\mathrm{Be}$ isotope are awaited, while a better understanding of the correlation properties of the AMS-02 uncertainties is desirable.

\section*{Acknowledgments}

We thank Jan Heisig and Tim Linden very helpful discussions and comments as well as for a carful reading of the manuscript.

The work of A.C. and M.K is supported by: ``Departments of Excellence 2018-2022'' grant awarded by the Italian Ministry of Education, 
University and Research (MIUR) L. 232/2016; Research grant ``The Dark Universe: A Synergic Multimessenger Approach'' 
No. 2017X7X85K, PRIN 2017, funded by MIUR; Research grant TAsP (Theoretical Astroparticle Physics) funded by INFN.

M.K further acknowledges support by Istituto Nazionale di Fisica Nucleare (INFN), 
by the Italian Space Agency through the ASI INFN agreement No.\ 2018-28-HH.0: "Partecipazione italiana al GAPS -- General AntiParticle Spectrometer", and the grant `The Anisotropic Dark Universe' No.\ CSTO161409, funded by Compagnia di Sanpaolo and University of Turin.
Furthermore, MK is partially supported by the Swedish National Space Agency under contract 117/19 and the European Research Council under grant 742104.

The majority of computations in this work were performed with computing resources granted by RWTH Aachen University under the project No. rwth0085.
Furthermore, computations were enabled by resources provided by the Swedish National Infrastructure for Computing (SNIC) under the project No. 2020/5-463
partially funded by the Swedish Research Council through grant agreement no. 2018-05973.

\bibliography{bibliography}{}
\bibliographystyle{apsrev4-1.bst}

\clearpage
\newpage

\appendix*
\section{Supplements}
\label{sec::app}

\begin{figure*}[b!]
\centering
\includegraphics[width=0.99\textwidth, trim={3cm 3cm 3cm 3cm},clip ]{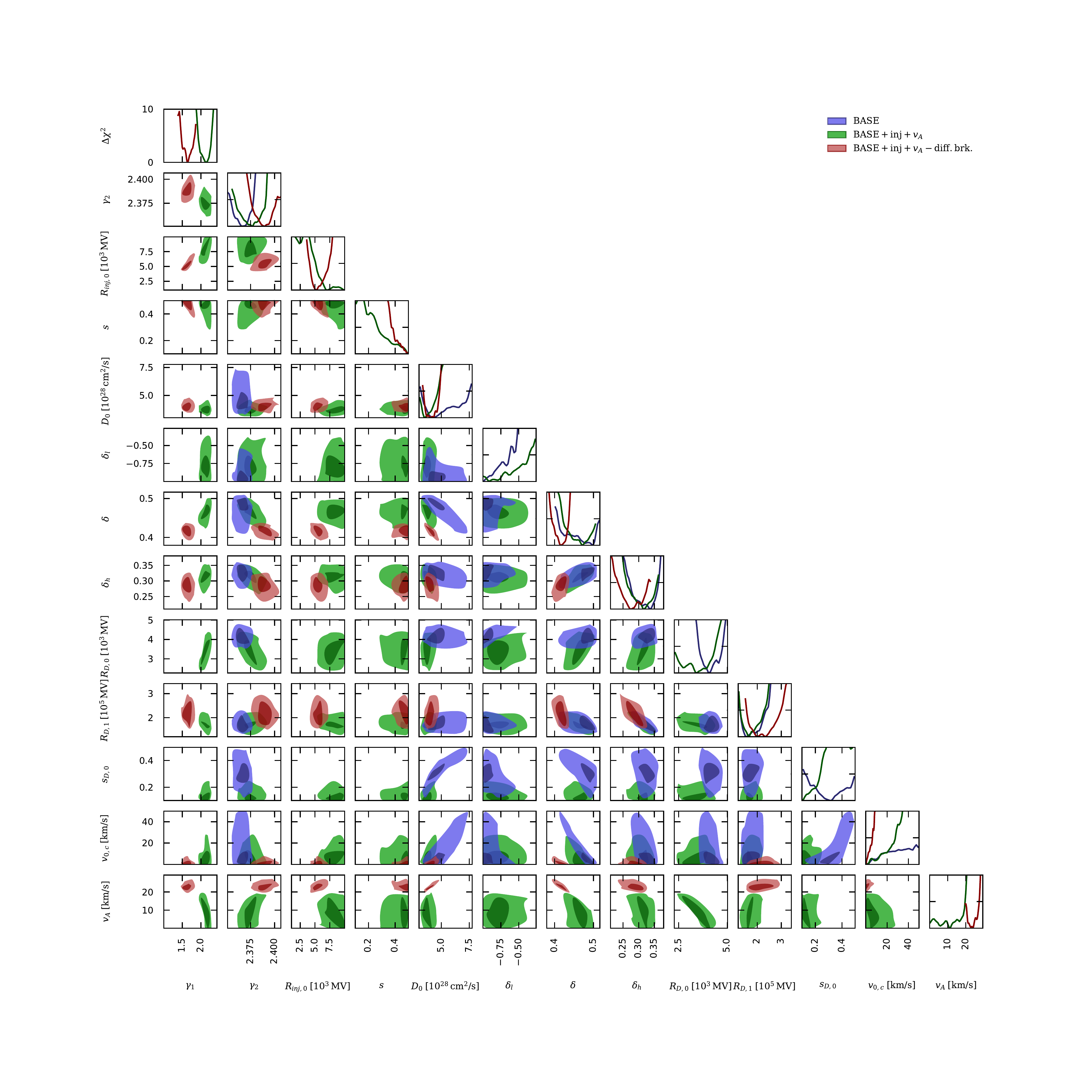}
\caption{ 
          Same as Fig.~\ref{fig:BCNO_propagation_triangle}, but for the fits with the LiBeBCNO data set.
          \label{fig:LiBeBCNO_propagation_triangle}
        }
\end{figure*}

\begin{figure*}[t!]
\centering
\includegraphics[width=1.0\textwidth, trim={3cm 3cm 3cm 3cm},clip ]{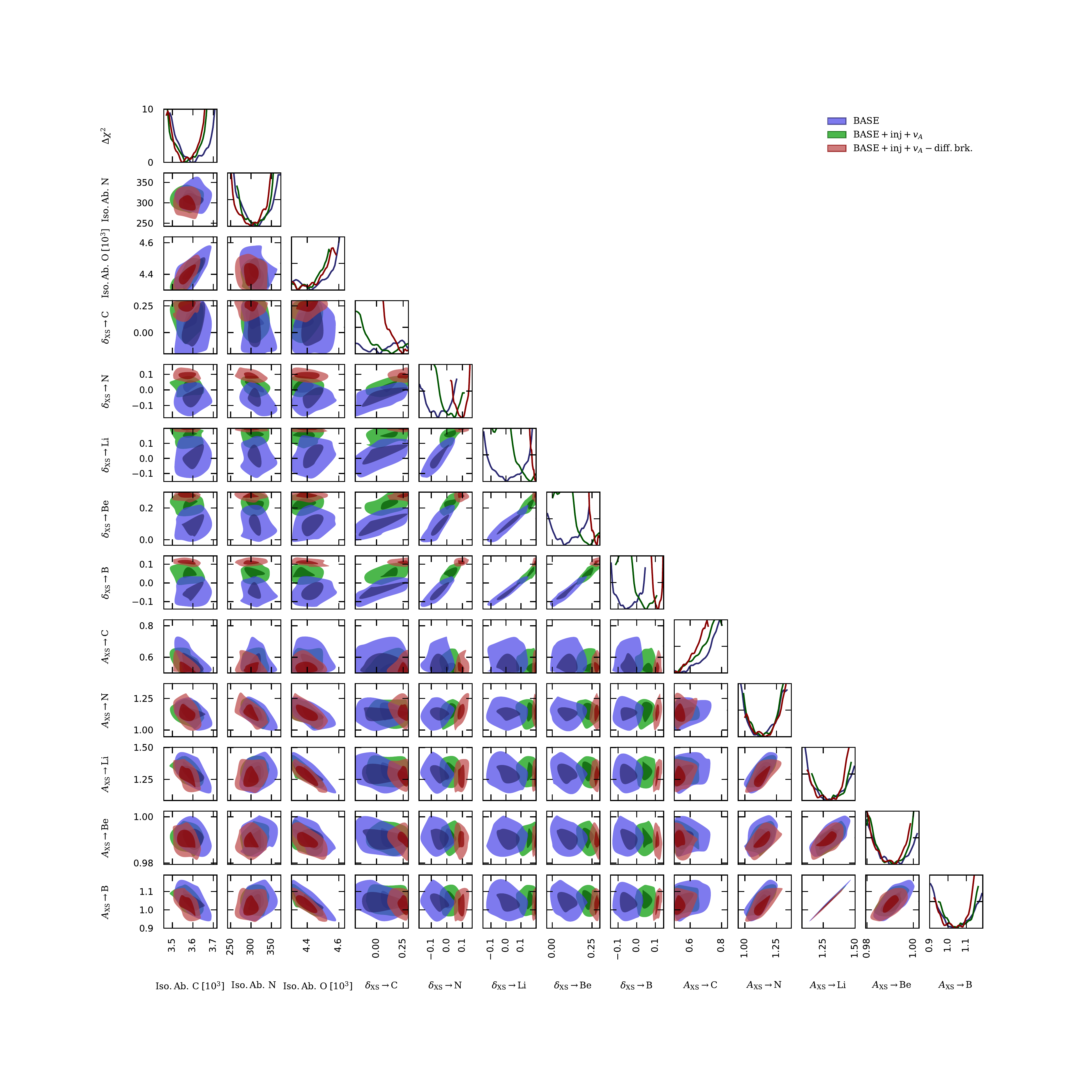}
\caption{ 
          Same as Fig.~\ref{fig:BCNO_propagation_triangle}, but for the fits with the LiBeBCNO data set, and for the subset of cross section nuisance parameters. 
          \label{fig:LiBeBCNO_XS_triangle}
        }
\end{figure*}

We list in this appendix further plots related to the main analysis described in the main text.
The first two, Figs.~\ref{fig:LiBeBCNO_propagation_triangle} and \ref{fig:LiBeBCNO_XS_triangle} are 
the equivalents of Figs.~\ref{fig:BCNO_propagation_triangle} and \ref{fig:BCNO_XS_triangle}, respectively, but for fits with the LiBeBCNO data sets. 
The results and conclusions are in agreement with those from the fits on the BCNO data set, as discussed in the main text.
We note that Figs.~\ref{fig:LiBeBCNO_XS_triangle} contains more cross section nuisance parameters compared to Figs.~\ref{fig:BCNO_XS_triangle}, namely, those 
related to the Be and Li spectra. As expected, we can observe that the slopes $\delta_\mathrm{XS}$ as well as the normalizations $A_\mathrm{XS}$ of Li, Be, and B are correlated.
Correlation is particularly pronounced between B and Li. This might possibly be due to the \textit{on-the-fly} sampling strategy applied to these two parameters. 

\begin{figure*}[t!]
\centering
\includegraphics[width=0.99\textwidth, trim={3cm 3cm 6.1cm 32.5cm},clip ]{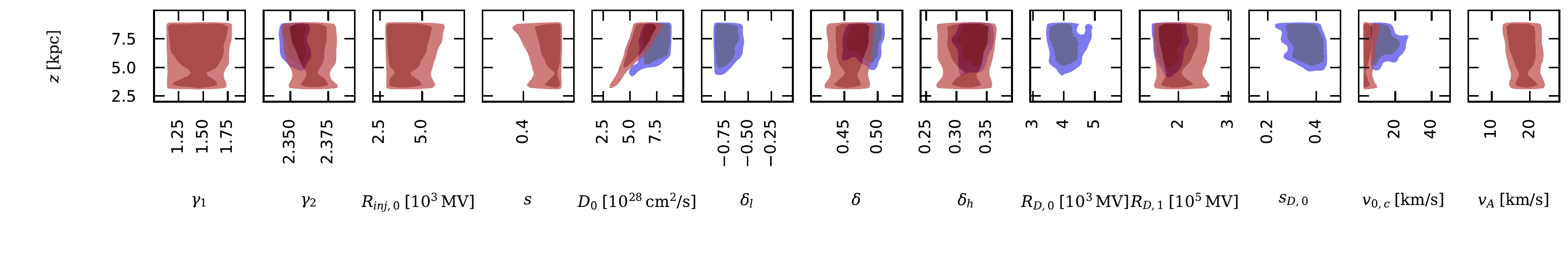} \vspace{1cm}\\
\includegraphics[width=0.99\textwidth, trim={3cm 3cm 6.1cm 32.5cm},clip ]{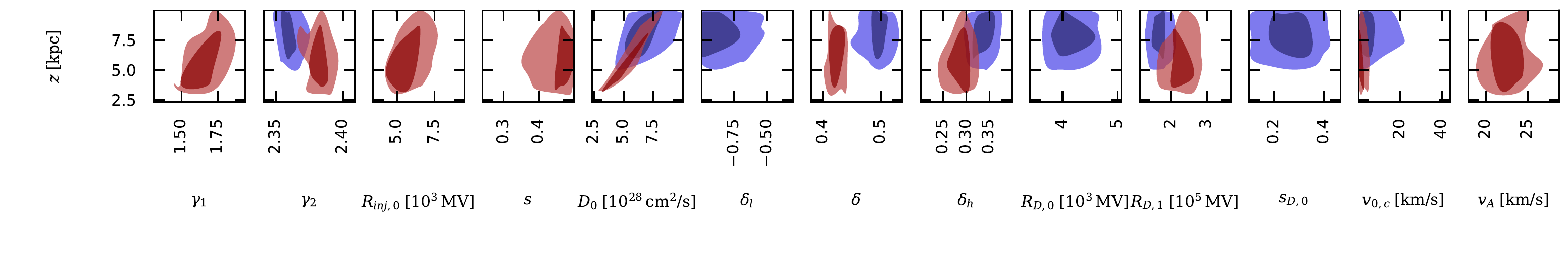}
\caption{ 
          Contours at the  1$\sigma$ and 2$\sigma$ C.L. showing the correlation between the halo half-height $z_\mathrm{h}$ 
          and all other CR propagation parameters for fits in the
          BASE (blue) and  BASE+inj+$v_A$$-$diff.brk (red) scenario. The upper panel corresponds to 
          the BCNO fits,  while the lower panel to the  LiBeBCNO fits. 
        }
\label{fig:zh_triangle}
\end{figure*}

\bigskip
Figure~\ref{fig:zh_triangle} contains additional results related to the fits which include the halo half-height $z_\mathrm{h}$ as free parameters (\textit{c.f.} Sec.~\ref{sec::zh}). 
We show the 2D uncertainty contours for all other CR propagation parameters with $z_\mathrm{h}$ for the BASE and BASE+inj+$v_A$$-$diff.brk setup. Results are shown both for the BCNO and LiBeBCNO data set. 
We note that the well-known degeneracy between the normalization of the diffusion coefficient $D_0$ and halo half-height $z_\mathrm{h}$ is 
less pronounced for the BASE framework. The reason for this is that in the BASE scenario the 
diffusion coefficient contains a break at around 4~GeV. Therefore,  $D_0$, which is normalized at a rigidity of 4~GV, is much more degenerate 
also with other parameters like $s_{D,0}$ and $\delta_l$. Consequently the correlation between $D_0$ and $z_\mathrm{h}$ is smeared out  in the BASE framework, while
while it remains well visible for BASE+inj+$v_A$$-$diff.brk one. 


\bigskip

\begin{figure*}[t!]
\centering
\includegraphics[width=1.0\textwidth, trim={3cm 3cm 3cm 3cm},clip ]{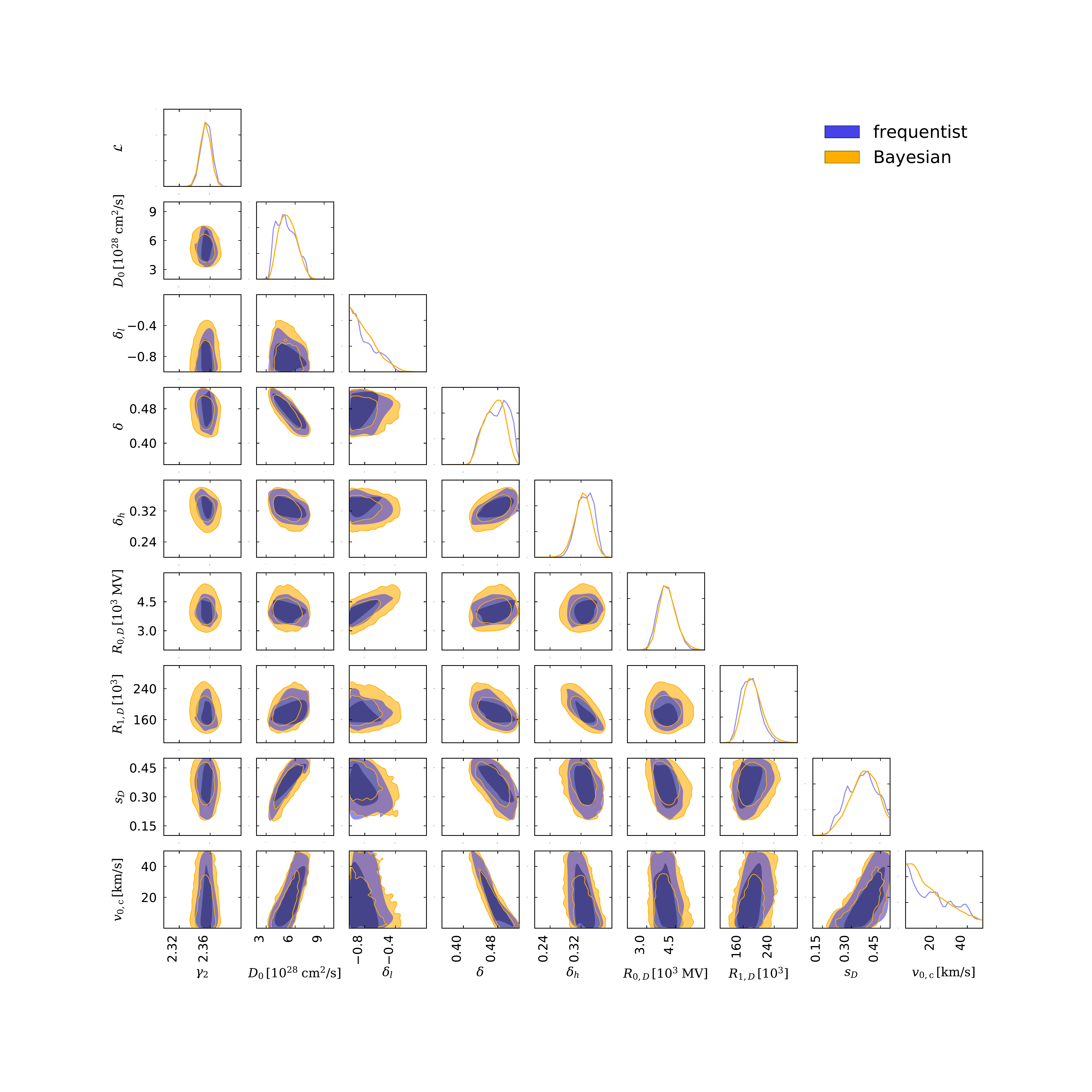}
\caption{ 
          Triangle plot with the fit results for the CR propagation frameworks  
          BASE and using the BCNO data set. We compare the frequentist (blue) and Bayesian (amber) 
          interpretation. The triangle plot shows only the subset of fit parameters related to CR propagation. 
          The 2D plots show the 1$\sigma$ and 2$\sigma$ contours for each combination of two parameters and the diagonal shows 
          the profiled likelihood and marginalized posterior for the frequentist and Bayesian interpretation, respectively. 
          \label{fig:Frequentist_vs_Bayesian}
        }
\end{figure*}

Finally, we show an example in which we compare the CR propagation parameters obtained in the Bayesian or frequentist statistical  interpretation,
namely the BASE framework of CR propagation and the fit to the BCNO data set. 
In Fig.~\ref{fig:Frequentist_vs_Bayesian} we compare the two statistical interpretation. For each couple of parameters the frequentist  
1 and 2 $\sigma$ contours (blue) are derived from the 2D $\chi^2$-profiles. On the other hand, the 
Bayesian 1 and 2 $\sigma$ contours (amber) are derived from the marginalized posterior function and correspond to the 68.3\% and 95.5\% credible level, respectively. 
On the diagonal we show the marginalized 1-dimensional posterior function for the Bayesian interpretation which is compared to the 1-dimensional profiled likelihood of the frequentist interpretation. 
As can be seen, the 1D profiled likelihood and 1D marginalized posterior on the diagonal of 
Fig.~\ref{fig:Frequentist_vs_Bayesian} match extremely well. We thus conclude that the constraints derived in this work do not depend on the statistical interpretation. 
We remark, however, that 2D Bayesian contours are slightly larger compared to the 2D frequentist contours meaning that the Bayesian interpretation is a bit more conservative.

\end{document}